\documentclass[final,5p,times,twocolumn,numbers]{elsarticle}
 
\usepackage[hyphenbreaks]{breakurl}
\usepackage{graphics}
\usepackage{adjustbox}
\usepackage{subcaption}
\usepackage{siunitx}
\usepackage[mathlines]{lineno}
\usepackage{textgreek}
\usepackage{multicol}
\usepackage{enumitem}
\usepackage{bm}
\usepackage{amsmath}
\usepackage{amssymb}
\usepackage{soul}
\usepackage{tabularx}
\usepackage{float}
\usepackage{booktabs}
\usepackage{array,multirow}
\usepackage[table]{xcolor}
\usepackage[T1]{fontenc}
\usepackage[hyperfootnotes=false]{hyperref}
\hypersetup{
		colorlinks   = true, %Colours links instead of ugly boxes
        linkcolor    = blue, %Colour of internal links
        citecolor = blue, %Colour of citations
		urlcolor = black,
		}

\let\origautoref\autoref
\def\autoref#1{{\origautoref{#1}}}

\definecolor{vanessa}{RGB}{0,0,0}
\definecolor{arsenii}{RGB}{0,0,0}
\definecolor{andrea}{RGB}{0,0,0}
\definecolor{comments}{RGB}{0,0,0}

\DeclareSIUnit{\MeV}{\mega\electronvolt}
\DeclareSIUnit{\keV}{\kilo\electronvolt}
\DeclareSIUnit{\us}{\micro\second}
\DeclareSIUnit{\ns}{\nano\second}
\DeclareSIUnit{\Hz}{\hertz}

\journal{Physics Letters B}

\begin{document}

\begin{frontmatter}

\title{Interpretable machine learning approach for electron antineutrino selection in a large liquid scintillator detector}

%% AUTHORS LIST

\author[padova-uni,padova-infn]{A. Gavrikov\fnref{em1}}
\fntext[em1]{arsenii.gavrikov@pd.infn.it}
\author[padova-uni,padova-infn]{V. Cerrone\fnref{em2}}
\fntext[em2]{vanessa.cerrone@pd.infn.it}
\author[padova-uni,padova-infn]{A. Serafini\fnref{em3}}
\fntext[em3]{andrea.serafini@pd.infn.it}

\author[padova-uni,padova-infn]{R. Brugnera}
\author[padova-uni,padova-infn]{A. Garfagnini}
\author[padova-uni,padova-infn]{M. Grassi}
\author[padova-uni,padova-infn]{B. Jelmini}
\author[padova-uni,padova-infn]{L. Lastrucci}

\author[catania]{S. Aiello}
\author[catania]{G. Andronico}
\author[milano]{V. Antonelli}
\author[bicocca]{A. Barresi}
\author[milano]{D. Basilico}
\author[milano]{M. Beretta}
\author[padova-uni,padova-infn]{A. Bergnoli}
\author[bicocca]{M. Borghesi}
\author[milano]{A. Brigatti}
\author[catania]{R. Bruno}
\author[roma]{A. Budano}
\author[milano]{B. Caccianiga}
\author[polimi]{A. Cammi}
\author[catania]{R. Caruso}
\author[bicocca]{D. Chiesa}
\author[perugia]{C. Clementi}
\author[padova-infn]{S. Dusini}
\author[roma]{A. Fabbri}
\author[frascati]{G. Felici}
\author[milano]{F. Ferraro}
\author[milano]{M. G. Giammarchi}
\author[catania]{N. Giudice}
\author[padova-uni]{R. M. Guizzetti}
\author[catania]{N. Guardone}
\author[milano]{C. Landini}
\author[padova-infn]{I. Lippi}
\author[roma]{S. Loffredo}
\author[polimi]{L. Loi}
\author[milano]{P. Lombardi}
\author[catania]{C. Lombardo}
\author[ferrara]{F. Mantovani}
\author[roma]{S. M. Mari}
\author[frascati]{A. Martini}
\author[milano]{L. Miramonti}
\author[ferrara]{M. Montuschi}
\author[bicocca]{M. Nastasi}
\author[roma]{D. Orestano}
\author[perugia]{F. Ortica}
\author[frascati]{A. Paoloni}
\author[milano]{E. Percalli}
\author[roma]{F. Petrucci}
\author[bicocca]{E. Previtali}
\author[milano]{G. Ranucci}
\author[milano]{A. C. Re}
\author[padova-uni,padova-infn]{M. Redchuck}
\author[ferrara]{B. Ricci}
\author[perugia]{A. Romani}
\author[milano]{P. Saggese}
\author[catania]{G. Sava}
\author[padova-uni,padova-infn]{C. Sirignano}
\author[bicocca]{M. Sisti}
\author[padova-infn]{L. Stanco}
\author[roma]{E. Stanescu Farilla}
\author[ferrara]{V. Strati}
\author[milano]{M. D. C. Torri}
\author[padova-uni,padova-infn]{A. Triossi}
\author[catania]{C. Tuvè}
\author[roma]{C. Venettacci}
\author[catania]{G. Verde}
\author[frascati]{L. Votano}

\affiliation[padova-uni]{organization={Università degli Studi di Padova, Dipartimento di Fisica e Astronomia}, country={Italy}}
\affiliation[padova-infn]{organization={INFN, Sezione di Padova}, country={Italy}}
\affiliation[bicocca]{organization={INFN, Sezione di Milano Bicocca e Dipartimento di Fisica Università di Milano - Bicocca}, country={Italy}}
\affiliation[polimi]{organization={INFN, Sezione di Milano Bicocca e Dipartimento di Energia, Politecnico di Milano}, country={Italy}}
\affiliation[catania]{organization={INFN, Sezione di Catania e Università degli Studi di Catania, Dipartimento di Fisica e Astronomia}, country={Italy}}
\affiliation[milano]{organization={INFN, Sezione di Milano e Università degli Studi di Milano, Dipartimento di Fisica}, country={Italy}}
\affiliation[perugia]{organization={INFN, Sezione di Perugia e Università degli Studi di Perugia, Dipartimento di Chimica, Biologia e Biotecnologie}, country={Italy}}
\affiliation[ferrara]{organization={INFN, Sezione di Ferrara, Università degli Studi di Ferrara, Dipartimento di Fisica e Scienze della Terra}, country={Italy}}
\affiliation[roma]{organization={INFN, Sezione di Roma Tre e Università degli Studi di Roma Tre, Dipartimento di Matematica e Fisica e Matematica}, country={Italy}}
\affiliation[frascati]{organization={INFN, Laboratori Nazionali dell'INFN di Frascati}, country={Italy}}

\begin{abstract}
Several neutrino detectors, KamLAND, Daya Bay, Double Chooz, RENO, and the forthcoming large-scale JUNO, rely on liquid scintillator to detect reactor antineutrino interactions. In this context, inverse beta decay represents the golden channel for antineutrino detection, providing a pair of correlated events, thus a strong experimental signature to distinguish the signal from a variety of backgrounds.
However, given the low cross-section of antineutrino interactions, the development of a powerful event selection algorithm becomes imperative to achieve effective discrimination between signal and backgrounds. In this study, we introduce a machine learning (ML) model to achieve this goal: a fully connected neural network as a powerful signal-background discriminator for a large liquid scintillator detector.
We demonstrate, using the JUNO detector as an example, that, despite the already high efficiency of a cut-based approach, the presented ML model can further improve the overall event selection efficiency. Moreover, it allows for the retention of signal events at the detector edges that would otherwise be rejected because of the overwhelming amount of background events in that region.
We also present the first interpretable analysis of the ML approach for event selection in reactor neutrino experiments.
This method provides insights into the decision-making process of the model and offers valuable information for improving and updating traditional event selection approaches. 
\end{abstract}


\end{frontmatter}

%% Main text

\section{Introduction}
\label{sec:intro}

Over the past decades, particle physics has experienced a paradigm shift in the data analysis approach, integrating well-established traditional methods with advanced machine
learning (ML) techniques~\cite{bib:ml-particlephysics, ml-schwartz}.
ML tools have found extensive applications in different fields of particle physics, including neutrino physics, offering solutions to a wide range of challenges, e.g., particle identification~\cite{bib:nova-pid}, background rejection~\cite{bib:ml-next}, vertex and energy reconstruction~\cite{bib:juno-rec, bib:juno-rec_en}, fast event generation \cite{bib:ratnikov_gans}, end-to-end detector optimization~\cite{bib:ratnikov_detectors}. {\color{vanessa} In neutrino physics, examples also include the use of deep convolutional neural networks for selecting inclusive charged-current interactions in MicroBooNE~\cite{bib:mb_example} and the adoption of boosted decision trees in Super-Kamiokande to improve multi-site tagging of electron neutrino-like events~\cite{bib:sk_example}.}
{\color{arsenii} Although these examples fall within the same scope as our study, it is difficult to make a quantitative comparison due to differences in the underlying principles of neutrino detection and/or the involved neutrino sources. However, a recent study on the application of machine learning methods for background rejection in KamLAND geo-neutrino analysis has similar experimental conditions and reports a significant improvement in the signal-to-background ratio~\cite{bib:km_example}.}

A common concern with the widespread adoption of ML methods is their perceived
\textit{black-box} nature. Many of these advanced algorithms lack
transparency, making it difficult for researchers to understand the
underlying mechanisms driving their predictions. This lack of
interpretability can be a significant barrier, especially in
scientific domains where a deep understanding of the processes
involved is crucial. In this study, we address this question by
focusing on the development of interpretable and explainable ML
methods~\cite{bib:ml-explainable, bib:bookinterpretable, bib:majorana} for the specific task of reactor antineutrino event selection in a liquid scintillator detector.\\
Reactor antineutrinos have held a crucial role in the neutrino physics landscape since their very first detection~\cite{bib:reines-cowan}. The most common channel for their detection is the Inverse Beta Decay (IBD) reaction, where an electron antineutrino interacts with a proton and produces a positron and a neutron, primarily due to its substantial cross section with respect to other processes in the MeV energy range~\cite{bib:vogel-xs}.
Many of the modern reactor experiments, such as KamLAND~\cite{bib:kamland}, Daya Bay~\cite{bib:dayabay}, Double Chooz~\cite{bib:doublechooz}, and RENO \cite{bib:RENO}, have adopted the Liquid Scintillator (LS) technology. This approach is based on using organic Hydrogen-rich materials that serve as both the proton target for the antineutrinos and the medium for detecting the outgoing positron. The resulting neutron can be captured by either isotopes present in the scintillator, such as Hydrogen itself or other elements like Carbon or Nitrogen, or by specifically loaded targets like gadolinium~\cite{bib:dayabay, bib:doublechooz,bib:RENO}.
We focus on the Jiangmen Underground Neutrino Observatory (JUNO)~\cite{bib:junophysics2016,bib:junophysics},
a multi-purpose and new generation LS experiment currently under construction in South China, largely exceeding its predecessors in size and expected performances. Its Central Detector (CD) is composed of a 20~kton liquid scintillator target housed within a 17.7-meter-radius spherical acrylic vessel and immersed in a 35~kton ultra-pure water pool. The CD is equipped with an advanced photo-detection system comprising 17612 20-inch photomultiplier tubes (PMTs) and 25600 3-inch PMTs, attached to a
surrounding Stainless Steel (SS) structure. This configuration provides an extensive total
photo-coverage of $\sim$78\%, granting a photoelectron (PE) statistics of $\sim$1600 PEs at 1 MeV~\cite{bib:nnn2023}.
\\
Due to the extremely small cross sections of neutrino weak interactions, neutrino events are inherently rare.  
For this reason, intensive efforts are dedicated to mitigating the backgrounds~\cite{bib:juno-radio}. Efficient control of radiogenic contamination
is achieved through meticulous detector design, careful selection of the employed materials, and strict radiopurity standards for the LS formula. Despite its underground location (1800 m.w.e.) and the expected high level of LS purification, the substantial size of the JUNO detector results in more significant background contamination than what is typically observed in smaller-scale experiments. As a result, performing an efficient event selection is of utmost importance in JUNO. \\
This paper is organized as follows: in \autoref{sec:benchmark_sel}, the IBD reaction mechanism and a cut-based benchmark selection strategy are presented, underlying the problem addressed with machine learning.{\color{arsenii}~\autoref{sec:problem_statement} introduces the ML models used in the study and discusses goals of the interpretability analysis of an ML model.}
In \autoref{sec:datadesc}, the data samples employed in the study are described. {\color{arsenii} In \autoref{sec:ml}, we discuss the machine learning approach in details, we describe how the ML models were trained and how their hyperparameters were optimized.} The performance of the presented ML models is discussed in \autoref{sec:results} and compared to the benchmark approach\footnote{Throughout this discussion, ``benchmark'' selection refers to the cut-based selection used in~\cite{bib:sub_osc}, which however does not necessarily reproduce JUNO official selection.}. \autoref{sec:xai} is dedicated to {\color{arsenii} the model's interpretability. \autoref{sec:mc_ue} discusses model calibration and uncertainty estimation of the model.} Finally, we conclude and summarize the work in \autoref{sec:conclusions}.

\section{Electron antineutrino detection and benchmark IBD selection}
\label{sec:benchmark_sel}

The selection rationale is driven by the characteristic
pattern yielded by the IBD reaction, where an antineutrino $\overline{\nu}^{}_e$ interacts with a proton in the Hydrogen-rich target medium, producing a positron and a neutron in the final state. The positron quickly deposits its
kinetic energy through ionization and annihilates with an electron into two \SI{511}{\keV}
photons, resulting in a \textit{prompt} signal. Meanwhile, the neutron
thermalizes in the detector and, after an average time of
\SI{220}{\micro\second}, undergoes capture on either Hydrogen or Carbon present in the LS. The subsequent
emission of a \SI{2.22}{\MeV}($\sim$99\% of cases) or \SI{4.95}{\MeV} ($\sim$1\% of cases) gammas, respectively, generates a \textit{delayed} signal.  It is worth mentioning that neutrons can be captured on other isotopes, like $^{13}$C, $^{14}$N, forming a delayed signal at higher energies in approximately 0.01\% of cases.

This double signature represents a powerful means to discriminate signal from backgrounds. The latter can be divided into two main groups. A correlated background consists of a pair of events induced by a single physics process, mimicking the prompt-delayed pattern induced by reactor antineutrino interactions (e.g., geoneutrinos and long-lived cosmogenic isotopes, $^9$Li and $^8$He, fall within this class~\cite{bib:junophysics}). 
On the other hand, uncorrelated backgrounds, often referred to as \textit{accidental coincidences}, arise when two independent signals are detected within a short time window, even though they are not associated with the same interaction (i.e., they mimic the typical time signature of signal events).
These coincidences are primarily attributed to radioactive contamination in the detector materials and surroundings. 
Some correlated backgrounds (e.g. geoneutrinos) are irreducible, others can be reduced through ad-hoc cuts (e.g. muon cuts for cosmogenic backgrounds), but their residual contamination is considered irreducible. Therefore the main task required of a selection algorithm is to distinguish between two classes: reactor antineutrino events and accidental coincidences.

As a benchmark for our ML model, we adapted the cut-based selection strategy from~\cite{bib:sub_osc}, which is based on a combination of cuts on the following six variables (or \textit{features}):
$E_{\rm prompt}$, $E_{\rm delayed}$, $R_{\rm prompt}$, $R_{\rm delayed}$, $\Delta t$, and $\Delta R$. The quantities $E_{\rm prompt}$, $E_{\rm delayed}$ represent the reconstructed energies, $R_{\rm prompt}$ and $R_{\rm delayed}$ are the radial components of the reconstructed vertices. Finally, $\Delta t$ is the time interval between prompt and delayed signals, and $\Delta R$ is the Euclidean distance between the two vertices.

All the aforementioned variables are obtained through a stand-alone Monte Carlo (MC) simulation based on the most up-to-date data published by the JUNO collaboration. Specifically, the generated data includes the simulation of the detector's geometry and the particles’ interactions inside the target material, e.g., physics processes such as light production and energy leakage, in order to produce data-like samples. 
Energies and vertices of the events are smeared with respect to their true values, following Ref.~\cite{bib:sub_osc}. During the real data taking and analysis of the experiment, variables will be provided either by JUNO reconstruction algorithms or by the official JUNO simulation software~\cite{bib:junosw} tuned on data. The cuts used for the benchmark approach are the following:
\begin{itemize}
    \item {\bf Fiducial volume (FV) cut:} prompt or delayed candidates are discarded if their vertices are reconstructed more than \SI{17.2}{\meter} away from the detector center. This cut is implemented to mitigate the impact of the exponentially increasing radioactive background rate at the edges of the target volume.
    \item {\bf Energy cut:} the energy windows are set as $E_{\rm prompt} \in (0.7, 12.0)$~MeV for prompt events and $E_{\rm delayed} \in (1.9, 2.5) \cup (4.4, 5.5)$~MeV for delayed events.
    \item {\bf Time cut:} the surviving pairs are required to fall in a time coincidence window of \SI{1}{\milli\second}, corresponding to approximately $5\times$ the neutron capture time.
    \item {\bf Vertex cut:} the distance between prompt and delayed events vertices has to be smaller than \SI{1.5}{\meter}, hence $\Delta R < \SI{1.5}{\meter}$.
\end{itemize}
Moreover, during data taking, an additional muon veto cut will be used, according to the topology of track-like events, resulting in a reduction of fiducial volume for a given time interval. The current state-of-the-art muon veto strategy~\cite{bib:sub_osc} yields a selection efficiency of 91.6\% for IBD events. We will not consider this criterion in our discussion, but it can be applied at a second stage as a multiplying factor for both ML and cut-based selection approaches.

\begin{figure*}[!htb]
	\centering
	\includegraphics[width=0.48\textwidth]{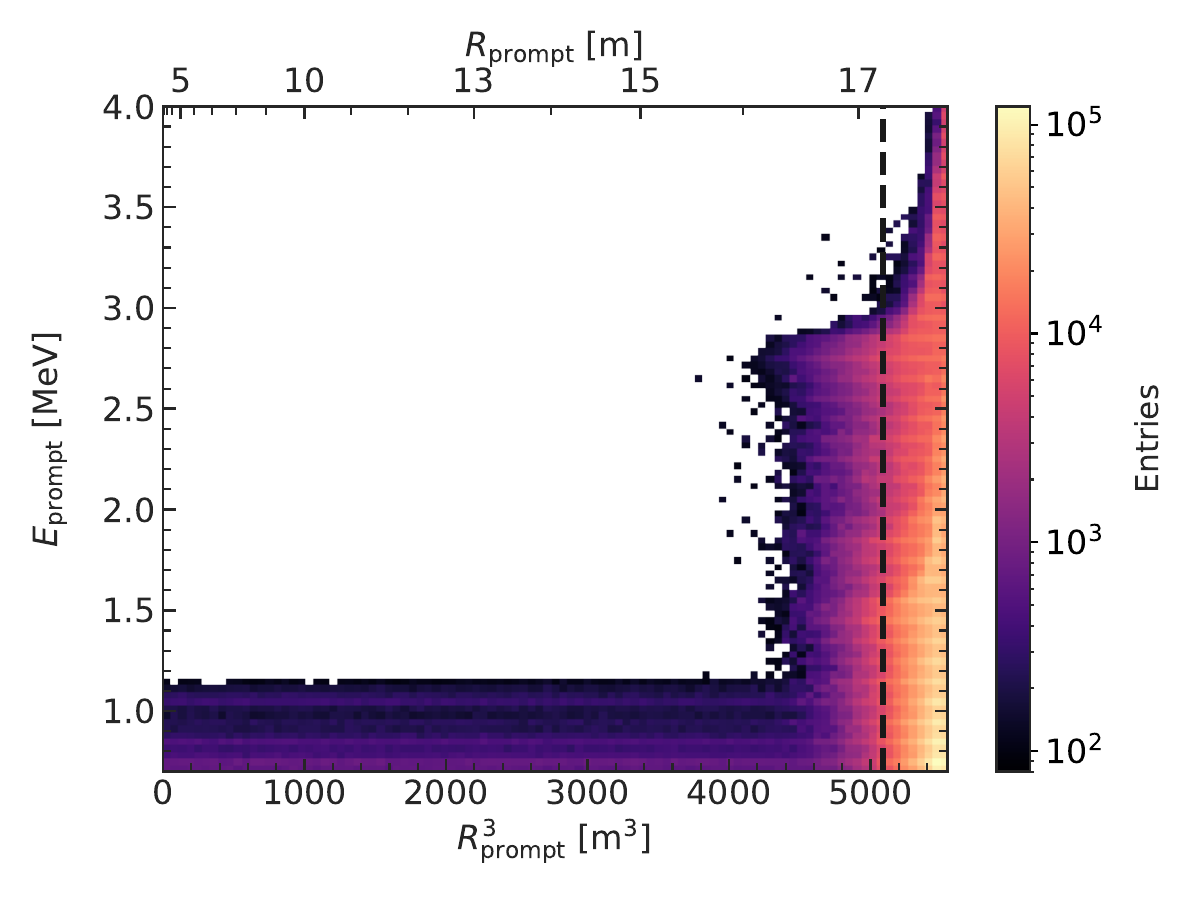}
	\hfill
	\includegraphics[width=0.48\textwidth]{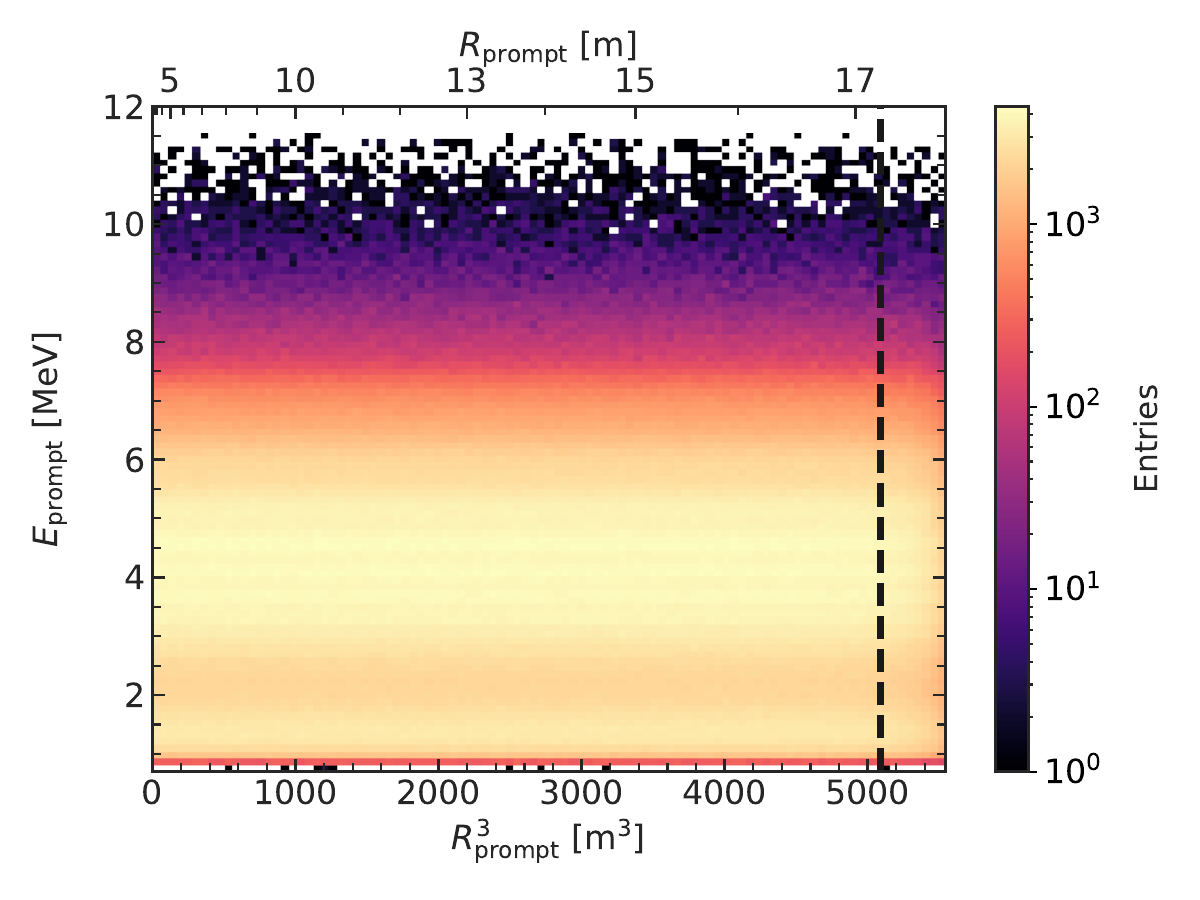}
	\caption{Reconstructed energy as a function of volume ($R^3$) for radioactivity (left) and IBD events (right). The IBD prompt energy spectrum extends up to approximately 12~MeV, while radiogenic events dominate the low energy range. The FV cut is indicated by the dashed line. The secondary axis provides the linear scale.}
	\label{fig:r3-energy}
\end{figure*}

Hereinafter we define \textit{efficiency} as the ratio between the amount of correctly classified IBD events and the total amount of IBDs in the dataset:
\begin{equation}
    \text{efficiency} = \frac{N_{\text{IBD}}^{\text{tagged as IBD}}}{N_{\rm IBD}}
    \label{eq:efficiency}
\end{equation}
This quantity corresponds, in the field
of ML, to the \textit{recall} metric,
measuring a classification model's
ability to correctly identify the positive class. 
On the other hand, \textit{purity} is associated with the residual background contamination and is defined as the ratio of correctly identified signal pairs to the total number of tagged as IBD events, represented by the equation:
\begin{equation}
    \text{purity} = \frac{N_{\text{IBD}}^{\text{tagged as IBD}}}{N^{\text{tagged as IBD}} }
    \label{eq:purity}
\end{equation}
The purity of a given sample is analogous to the ML \textit{precision}, gauging how accurately a machine learning model predicts the positive class. 
The selection efficiency for each cut is determined by calculating the ratio of the number of events meeting the specific criterion to the total number of reconstructed events before the application of that particular cut. This step-wise application of cuts is viable for IBD events due to the almost uncorrelated nature of all features. In contrast, accidental coincidences require a simultaneous application of all selection criteria to capture the intrinsic (and significant) dependence among features. An example is reported in \autoref{fig:r3-energy}.
(left for radiogenic events, right for IBD prompt candidates), where the event rate is shown as a function of $R^3$ and reconstructed energy.
IBD events are uniformly distributed inside the CD since antineutrinos are homogeneously interacting in the detector.
Contrariwise, a strong correlation exists between energy and radial distributions for radioactivity events. As a consequence, selection efficiency terms cannot be computed independently and then progressively combined.
In the subsequent sections, we will outline how this challenge can be effectively tackled using ML techniques.

The FV cut results in a sensitive loss of statistics for IBD events, of the order of 8\%, as it can be determined by  strictly geometrical considerations. Thus, one of the goals of using a ML algorithm is to create a more flexible demarcation between the two classes of events.
This flexibility would allow us to retain a greater number of signal events, maintaining the same, or potentially improved, purity level.
Furthermore, we observe that the energy cut is effective in rejecting background events and preserves nearly all IBD candidates, while still offering limited room for improvement. Furthermore, the conventional ``box-like" cut applied to the $\Delta R$ and $\Delta t$ features ($\Delta R < 1.5$ m, $\Delta t < 1$ ms) is sub-optimal, and analytical optimization becomes challenging when dealing with multi-dimensional PDFs. The single impact of the cuts on efficiency is discussed in detail in~\autoref{sec:ml_vs_standard}.

In addition, the accidental coincidences ($R^3$, energy) bi-dimensional distribution in \autoref{fig:r3-energy} (left panel) suggests that cuts that depend on the vertices distance $\Delta R$ and energy can potentially help in distinguishing between the two classes. Indeed, IBD candidates are to a large extent uniformly distributed while background events show a non-trivial radial distribution at different energies.
These specific examples underscore the potential of explainable ML techniques to (1) identify optimal cuts and (2) offer valuable insights into the relationships between features, empowering the analyzers with the capability to make informed decisions.

\section{Problem statement}
\label{sec:problem_statement}

As explained in the preceding section, the benchmark selection strategy is based on the use of relatively basic cuts, which, while effective, fail to address the inherent correlations in high-dimensional data. 
On the other hand, machine learning methods are proven to be powerful tools to process high-dimensional data and to find underlying non-linear dependencies within it.
{\color{arsenii} In this study, we use a fully connected neural network (FCNN) and Boosted Decision Trees (BDT)~\cite{bib:bdt1, bib:bdt2} as classifiers to distinguish between signal (reactor antineutrino) and background (random coincidence) events. BDT is a well-established approach for handling low-dimensional, high-level, tabular data~\cite{bib:bdt_vs_nn} and we use it as a baseline ML model to serve as comparison with the neural network. One limitation of tree-based methods is that, due to the nature of the algorithms, the decision boundaries between classes are represented by non-smooth, step-like functions. While BDT can still be highly effective in separating signal from background, this characteristic makes it challenging to directly apply the learned decision profiles as optimized cuts for a cut-based selection approach. In contrast, FCNN naturally provides smooth and differentiable boundaries between the classes, making neural networks a more desirable and general approach for achieving the objectives of this study. The classifiers use} as input the following ten features, complementing the cut-based selection set with angular information: $E_{\rm prompt}$, $E_{\rm delayed}$, $R^3_{\rm prompt}$, $R^3_{\rm delayed}$, $\cos(\theta_{\rm prompt})$, $\cos(\theta_{\rm delayed})$, $\varphi_{\rm prompt}$, $\varphi_{\rm delayed}$, $\Delta R$, and $\Delta t$. Here, $\varphi_{\rm prompt}$ and $\varphi_{\rm delayed}$ are the azimuthal angles, $\theta_{\rm prompt}$ and $\theta_{\rm delayed}$ the zenith angles with respect to the vertical $z$ axis. The choice of this particular set of features is driven by both the geometrical structure of the detector and the unique patterns observed in signal events. In brief, the variables $R^3$, $\varphi$, and $\cos \theta$ exhibit a uniform distribution for IBD candidates, whereas a more complex trend is expected for radiogenic background events. A detailed explanation of this particular aspect will be provided in \autoref{sec:datadesc}.
We acknowledge that using the prompt energy as an input feature for the model causes energy-dependent efficiency and purity estimations and this has to be properly taken into account at the level of subsequent analyses.

{\color{arsenii} A central part of the study is a comprehensive analysis of an ML model's interpretability.} This analysis has several goals:
\begin{enumerate}
\item Ensure trust in the model and its transparency by a deep understanding of the dependencies between features.
\item Achieve an understanding of decision boundaries between different classes, that can be provided by interpretable ML models in both visual and quantitative ways.
\item Optimize and fine-tune the cut-based selection criteria. An estimation of the importance of each feature within the selection task both at the level of the entire dataset and at the level of each individual event can potentially help in improving the efficiency of traditional event classification. {\color{arsenii} This can be especially important during the initial phase of data-taking since event generators and reconstruction algorithms may not perform optimally. As a result, one can expect noticeable discrepancies between Monte Carlo simulations and real data. Training an ML model on inaccurate simulations could lead to unreliable performances, as the robustness of the algorithm to this discrepancy remains uncertain.}

\end{enumerate}

{\color{arsenii} In the JUNO experiment, the tuning of detector response in the Monte Carlo simulation and the validation of reconstruction algorithms is planned to be performed using calibration data~\cite{bib:calibration_of_juno, bib:neuro_mct}. Once this tuning is completed, ML models can be retrained to achieve more reliable performance.}

\section{Data Description}
\label{sec:datadesc}

\begin{figure*}[!htb]
	\centering
	\includegraphics[width=1\textwidth]{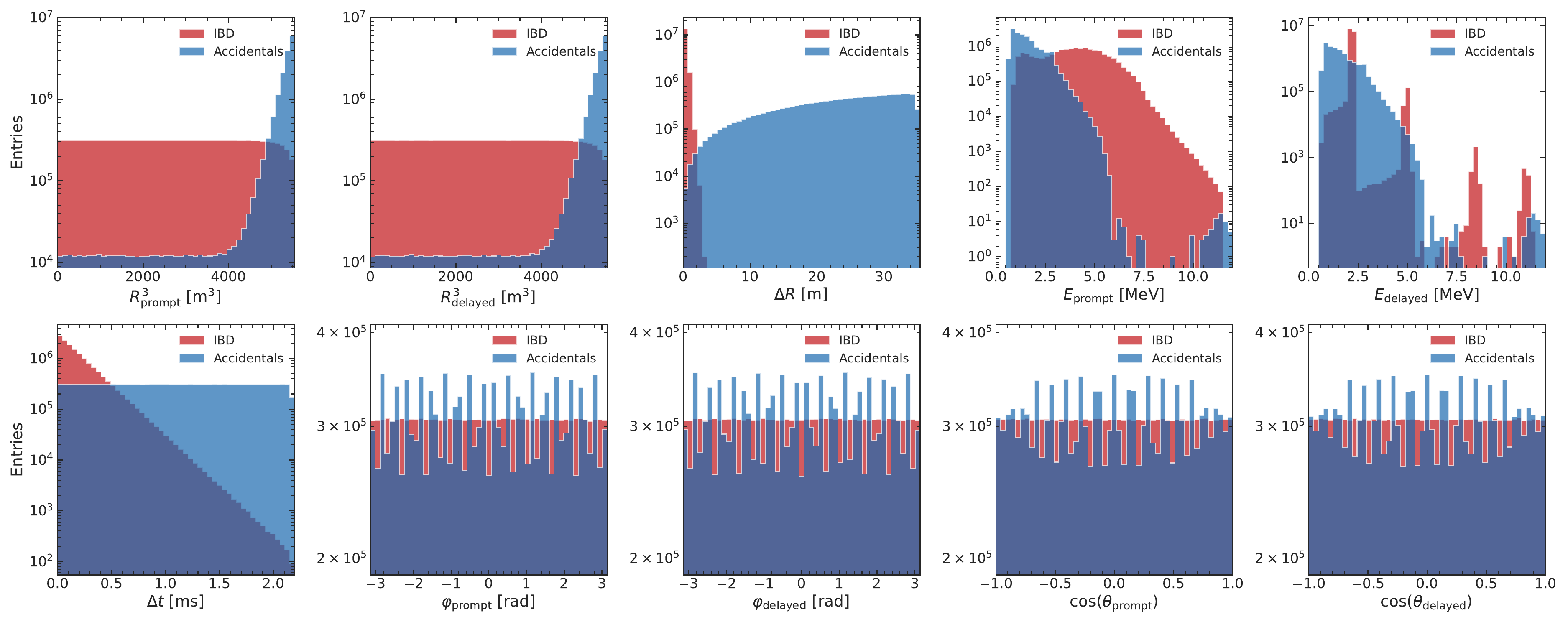}
	\caption{Distributions of features for the dataset used to train and evaluate the ML model for both IBD (red) and accidental (blue) events. }
	\label{fig:feature_distributions}
\end{figure*}

In this study, a standalone MC is used to generate data-like samples to test our selection strategy.
In particular, two different datasets were prepared:
\begin{enumerate}
    \item \textbf{IBD dataset}: it consists of 15M independent IBD pairs uniformly distributed in the full CD volume with radii up to 17.7~m. The energy distribution follows the expected oscillated spectrum of reactor electron antineutrinos.  It is worth mentioning that the two mass ordering assumptions are equivalent for our purpose, hence we chose the normal ordering current global best fit values to build out our dataset~\cite{bib:nufit}.
    \item \textbf{Accidentals dataset}: it consists of 15M pairs of different radioactive decays of all types, namely $\alpha, \beta, \gamma$~\cite{bib:juno-radio}. 
    To prevent biases and ensure the model's generalization capability, the same amount of events is chosen to balance the IBD dataset. This number of accidental events corresponds to approximately 50 days of data collection. The radioactivity of the materials used in the construction of the detector represents one of the main sources of accidental background. These radioactive contaminants release energy through their decay processes, and they are categorized as \textit{internal} if they are produced in the LS, or \textit{external}~\cite{bib:juno-radio} if they arise from other components of the detector, respectively. 
    \textit{Internal} radioactive events are uniformly distributed in the full CD volume with radii up to 17.7~m. \textit{External} radioactive events are instead generated at the detector edges and radially decrease following an exponential distribution going towards the detector center, because of their interaction with the LS itself \cite{bib:juno-radio}. While the \textit{internal} radioactivity is simulated as latitudinally and longitudinally homogeneous, the \textit{external} contribution is simulated with an angular modulation due to the grid structure of the detector components. Both components contribute to the resulting accidentals dataset's features.
  
\end{enumerate}

\begin{table}[!htb]
\renewcommand*{\arraystretch}{1.25}
    \centering
    \begin{tabular}{|c|c|}
        \hline
        \textit{Location} & \textit{Isotopes} \\
        \hline
        LS & $^{238}$U, $^{232}$Th, $^{40}$K, $^{210}$Pb, $^{14}$C, $^{85}$Kr \\
        \hline
        Acrylic & $^{238}$U, $^{232}$Th, $^{40}$K \\
        \hline
        SS & $^{238}$U, $^{232}$Th, $^{40}$K, $^{60}$Co \\
        \hline
        Glass & $^{238}$U, $^{232}$Th, $^{40}$K, $^{208}$Tl  \\
        \hline
        Water pool & $^{222}$Rn \\
        \hline
    \end{tabular}
    \caption{ Main radioactive contaminants and corresponding sources~\cite{bib:juno-radio}: 1) LS, 2) acrylic sphere, 3) stainless steel (SS) structure, 4) PMTs glass, 5) water pool.}
    \label{tab:acc_dataset}
\end{table}

\subsection{Data preparation}
To build the feature table for both datasets, we iterate over all events in the sample, considering them as prompt candidates. Then, for each $i$-th prompt event, we select all $j$-th events (with $i<j$) occurring in a time window of $10~\tau$, where $\tau~\simeq~\SI{220}{\micro\second}$ is the mean neutron capture time. This particular choice is aimed at minimizing event loss, ensuring that the fraction of potential candidates to be excluded is less than \num{5e-5}. Finally, we compute and store the relevant features for all possible $(i,j)$ combinations within the specified time interval\footnote{\color{arsenii} The average number of combinations per event is approximately 0.5 for IBDs and approximately 0.1 for accidentals. We would like to emphasize that the 0.5 value for IBDs arises because, for each prompt event, a delayed event can always be found within $10\tau$, but the delayed event does not have a subsequent event to form a pair within the window, taking into account the very low event rate of IBD.}.
Afterward, the two feature tables are merged, assigning the corresponding class (IBD or accidental) to each event.~\autoref{fig:feature_distributions} shows the distributions of the 10 features after all steps described above, for both IBD events (red) and accidental coincidences (blue).

 Hereinafter, we analyze the feature distributions in detail:
\begin{itemize}
    \item Accidental coincidences are not uniformly distributed within the LS and exhibit an exponential increase towards the edges of the detector, as expected. At the same time, it is worth noting that their radial coordinate (i.e., $R_{\rm prompt}^3$ and $R_{\rm delayed}^3$ in blue in the first two panels of \autoref{fig:feature_distributions}) is approximately uniform in the target volume up to about \SI{16}{\meter} ($\simeq$~\SI{4000}{\cubic\meter}).
    IBD events are instead uniformly distributed inside the CD, as previously mentioned. 

    \item The Euclidean distance $\Delta R$ between IBD prompt-delayed candidates' vertices is peaked at approximately $\sim$0.2~meters and the distribution depends on the random walk process of the emitted neutrons. As for random coincidences, the $\Delta R$ distribution is shaped by the spatial distribution of radiogenic events within the CD. 
    \item The energy distribution corresponds to the
    positron spectrum, and
    to the gammas emitted by neutron capture, for prompt and delayed IBD candidates, respectively.
    On the other hand, the radioactive decays of the primary contaminants determine the energy spectrum shape for accidental coincidences \cite{bib:juno-radio}, which is the same for both prompt and delayed events, except for statistical fluctuations. This spectrum has a prominent peak at energies $\sim$\SI{1}{\MeV}, where the major contribution comes from $^{14}$C and quenched $\alpha$ peaks, mainly from the $^{238}$U/$^{210}$Pb chains~\cite{bib:juno-radio}, and it extends up to \SI{5}{\MeV}, at the end point of $^{208}$Tl $\beta$ decay.
    \item The $\Delta t$ distribution for IBD events is an exponential decay with characteristic time related to neutron capture, while it is almost flat for accidentals. Specifically, the expected distribution is exponential with a long half-life determined by the event rate. 
    \item IBD events exhibit a spherical symmetry, resulting in a uniform distribution for $\varphi$ and $\cos\theta$. In contrast, radioactivity presents a distinctive non-uniformity due to contamination from the detector supporting structure,
    which are localized at fixed positions in $\varphi$ and $\cos\theta$. This deviation from uniformity is noticeable at the edges of the detector: this effect can be seen in \autoref{fig:feature_distributions}. 
\end{itemize}

\section{Machine Learning Approach}
\label{sec:ml}

In the context of machine learning, our goal --- selection of IBD events among accidental background --- is a supervised classification problem. In supervised learning problems, a model considers input-target pairs and learns the mapping from input features to a target value (or so-called \textit{label}). This learning process is based on using data samples with known input-target pairs. Depending on the type of target, one can define two types of supervised problems: classification problem (the target represents a discrete set of values) or regression problem (the target represents continuous values). More formally, let us have a set of pairs: $(\mathbf{x}_1, y_1), (\mathbf{x}_2, y_2), ..., (\mathbf{x}_n, y_n) = \{\mathbf{x}_i, y_i\}_{i=1, ...,n}$, where $\mathbf{x}_i \in \mathbb{R}^p$, $y_i \in \{0, 1\}$, $p$ is the number of input features, and $n$ is the amount of events in a data sample. The mapping from $\mathbf{x}$ to $y$ is then defined by a function $h: y = h(\mathbf{x})$. Our classification task is then to find a model $f:\mathbb{R}^p \longrightarrow \mathbb{R}$ which is a function of both $\mathbf{x}$ and parameters $\mathbf{\phi}$. The set of parameters $\mathbf{\phi}$ specify a relationship between an input and an output of a model. The discrepancy between the output and the target, between $f$ and $h$, is quantified with a function called \textit{loss} $L$. Training a model means to find a set of $\mathbf{\phi}$ which make the model $f$ best approximate the function $h$, so minimizes the loss function: $\hat{\mathbf{\phi}} = \underset{\mathbf{\phi}}{\mathrm{argmin}} [ L(\mathbf{\phi})]$.

The set of pairs $\{\mathbf{x}_i, y_i\}_{i=1, ...,n}$ is called \textit{training} dataset, i.e. the one used to directly train a model. Usually, to perform proper training and model evaluation procedures, one needs two additional datasets: a \textit{validation} dataset and a \textit{testing} dataset. The former is used to optimize the hyperparameters~\footnote{Hyperparameters --- parameters of a model that define its structure and its learning process. Hyperparameters are set before training is started and cannot be adjusted during the training, unlike learnable parameters (e.g. weights in a neural network). An example of hyperparameters of a neural network could be the following parameters: number of layers, number of units in a layer, learning rate, etc.} of a model and to evaluate the performance \textit{during} the training process. Conversely, the latter is used to test the performance of a model once it is \textit{trained}. For our task, the 30M dataset is split into three parts with the following ratios: 20M events for training, 5M events for validation, and 5M events for testing. This choice is made to have enough data for training, tuning hyperparameters, and evaluating the final model's performance, while ensuring that each subset is representative of the overall dataset.

{\color{arsenii} In this study, FCNN and BDT are used as a model $f$.}

\subsection{Boosted decision trees}
{\color{arsenii} BDT is a gradient boosting-based algorithm that combines multiple weak decision trees into a single strong model. The ensemble of BDT trees is built sequentially: each new tree is designed to correct the predictions made by the ensemble in the previous step. To avoid overfitting, trees usually have a relatively small maximum depth of less than 5. The boosting process reduces both bias and variance, improving overall prediction performance. As mentioned in \autoref{sec:problem_statement}, BDT is a powerful approach for low-dimensional, high-level, tabular input data and we use it as a baseline model for comparison with the FCNN. 

Additionally, one of the advantages of BDT is its ability to provide an independent way to access global feature importance through the algorithm’s construction. One method for this is given by the gain feature importance~\cite{bib:xgboost}, which is calculated as the average reduction in the loss function when an input feature is used to split a node in a tree. We use this metric as an additional cross-check for the feature importances computed for the FCNN model.

In this paper, we use the XGBoost~\cite{bib:xgboost} implementation of BDT as a robust and widely accepted framework for adopting the algorithm. To optimize the hyperparameters, we use the Random Sampler from the Optuna library~\cite{bib:optuna}. The total number of trees is determined through the early stopping procedure with a \textit{patience} parameter of 20, which indicates the maximum number of trees that can be added to the ensemble without any improvement on the validation dataset before the training process is stopped. The binary cross-entropy loss is used as the loss function. The optimized BDT model, with a maximum depth of \textbf{2}, a learning rate of \textbf{0.25}, and a total of \textbf{483} trees, is used further in the study.}

\subsection{Fully connected deep neural network
}

\begin{figure}[!htb]
	\centering
	\includegraphics[width=0.8\columnwidth]{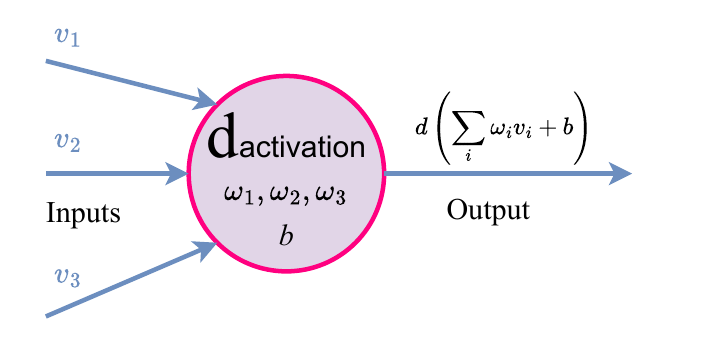}
	\caption{The schematic view of a neuron --- the basic component of a neural network.}
	\label{fig:neuron}
\end{figure}

\begin{table*}[!htb]
\renewcommand*{\arraystretch}{1.25}
	\centering
    \begin{tabular}{|l@{\hskip 60pt}|l|l@{\hskip 40pt}|}
    \hline
        \it{Hyperparameter} & \it{Search space} and \textbf{selected hyperparameter} \\
    \hline
	Units in input layer & [16, 256]: \textbf{96} \\
    \hline
	Units in hidden layers & [16, 256]: \textbf{240} \\    \hline
	Number of hidden layers & [1, 10]: \textbf{2} \\
    \hline Activation~\cite{bib:relu, bib:lrelu, bib:silu, bib:prelu} & \textbf{ReLU}, Leaky ReLU, SiLU, PReLU, Tanh \\
    \hline
	Optimizer~\cite{bib:adam, bib:sgd, bib:rmsprop} & \textbf{Adam}, SGD, RMSprop \\
    \hline
	Learning rate & [$10^{-5}$, $10^{-1}$]: $\mathbf{3.5 \cdot 10^{-4}}$ \\
    \hline
    Scheduler type~\cite{bib:ExpScheduler, bib:CosAnnScheduler} & Exponential, ReduceOnPlateau, \textbf{CosineAnnealing}, None \\
    \hline
	Layer weights initialization~\cite{bib:xavier, bib:orthogonal} & xavier uniform, xavier normal, \textbf{orthogonal}, normal, uniform \\
    \hline
	Batch normalization~\cite{bib:batchnorm} & True, \textbf{False} \\
    \hline
	Batch size & [128, 2048]: \textbf{1024} \\
    \hline
	\end{tabular}
	\caption{Hyperparameter search space for FCNN. Selected hyperparameters are highlighted in bold.}
	\label{tab:hypspace}
\end{table*}

~\autoref{fig:neuron} shows the basic component of an FCNN, i.e., a \textit{neuron} (or a \textit{unit}). Neurons are connected with other neurons and the strength of their connection is defined by weights $\omega_i$. These weights are adjusted during the training process to minimize the difference between the predicted and true outputs. Each neuron computes a weighted sum of inputs and then applies an \textit{activation function}:
\begin{align*}
    g(\textbf{v}) = d \left(\sum_i \omega_i v_i + b \right),
\end{align*}
where $b$ is a bias, $v_i$ are the inputs (usually, outputs of neurons of the previous layer, or the values of features in the case of the first layer), $d$ is an activation function, and $g$ is a neuron output. In order to build an FCNN model that is able to reproduce complex nonlinear dependencies in the data, the activation functions in the neurons must be nonlinear. Otherwise, in the case of their linearity, the entire neural network could be reduced to a linear mapping. There are many different nonlinear activation functions and more details about them can be found in Ref.~\cite{bib:activation_functions}.

\begin{figure}[!htb]
	\centering
	\includegraphics[width=1\columnwidth]{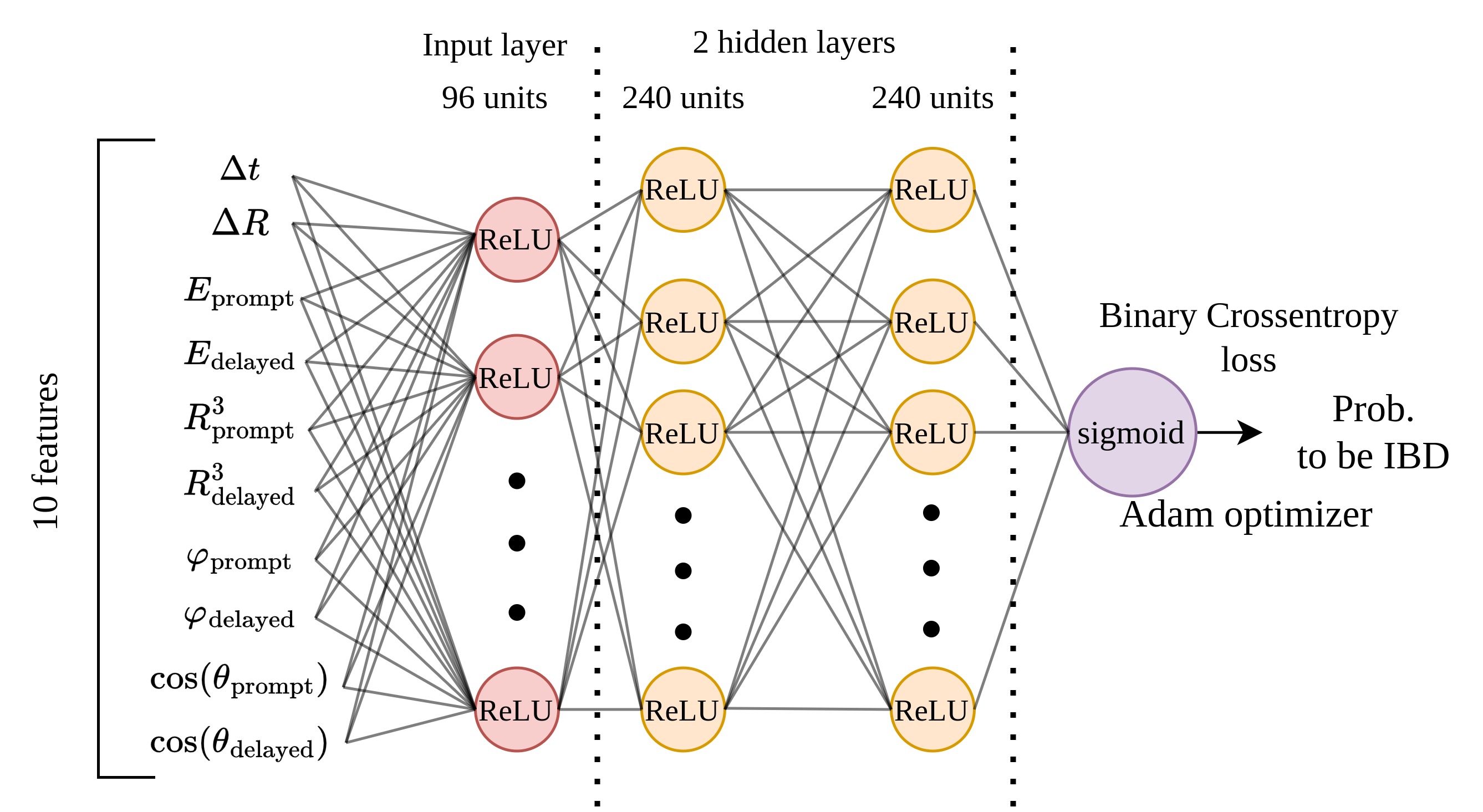}
	\caption{Network architecture after the optimization procedure. The 10 features introduced in~\autoref{sec:problem_statement} are used as input for a fully connected neural network with 3 layers: the input layer with 96 neurons and 2 hidden layers of 240 neurons. As an activation function for the neurons, we use ReLU functions for all the layers except for the output one where with the sigmoid function is used. Binary cross-entropy~\cite{bib:logloss} is used as a loss function, and Adam is used as an optimizer. The model consists of 84k trainable parameters. Being small and compact, the model can provide predictions for more than 1M events per second.}
	\label{fig:fcnn}
\end{figure}

In an FCNN, neurons are organized into layers, where each neuron in the layer is connected to all neurons from the previous one. Such a neural network can be divided into three main parts: the input layer, the hidden layers, and the output layer. The input layer receives features that describe a physical event, while the output layer gives the prediction of the model (in our case, the classification score from 0 to 1).
Hidden layers allow the model to expand the space of functions that it is able to approximate. 
A wide variety of hyperparameters define a neural network and their optimization is an important part of building the final model. In this study, hyperparameter optimization is performed using the Tree-structured Parzen Estimator algorithm~\cite{bib:tpe} from the Optuna library.~\autoref{tab:hypspace} shows the hyperparameters search space and the selected hyperparameters. We use the PyTorch framework~\cite{bib:torch} to build and train the model. It takes approximately two days to perform hyperparameter optimization and one hour to train the selected model on a Nvidia A30 GPU.
The binary cross-entropy loss~\cite{bib:logloss} is used as a loss function and the sigmoid~\cite{bib:activation_functions} is used as an activation function for the output layer. All input features were normalized with a standard score normalization. The training process is performed with an early stopping condition on the validation dataset with a \textit{patience} of 20. Here, the patience parameter refers to the number of epochs the training process is allowed to continue without any improvement (on the validation dataset) before being stopped.
~\autoref{fig:fcnn} shows the optimized FCNN architecture and its main hyperparameters.

\begin{figure*}[!htb]
	\centering
	\includegraphics[width=0.48\textwidth]{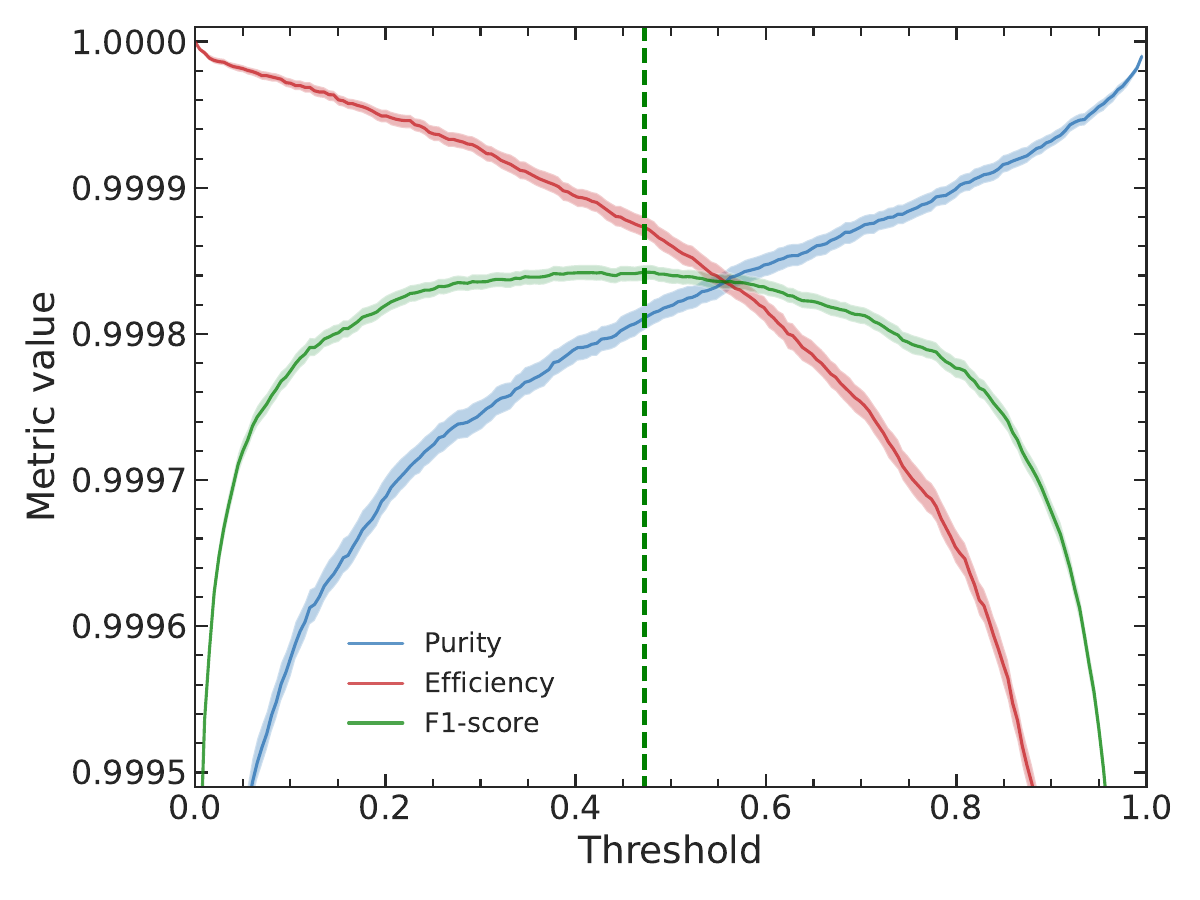}
% 	\label{fig:fcnn_output_threshold}
	\hfill
	\includegraphics[width=0.48\textwidth]{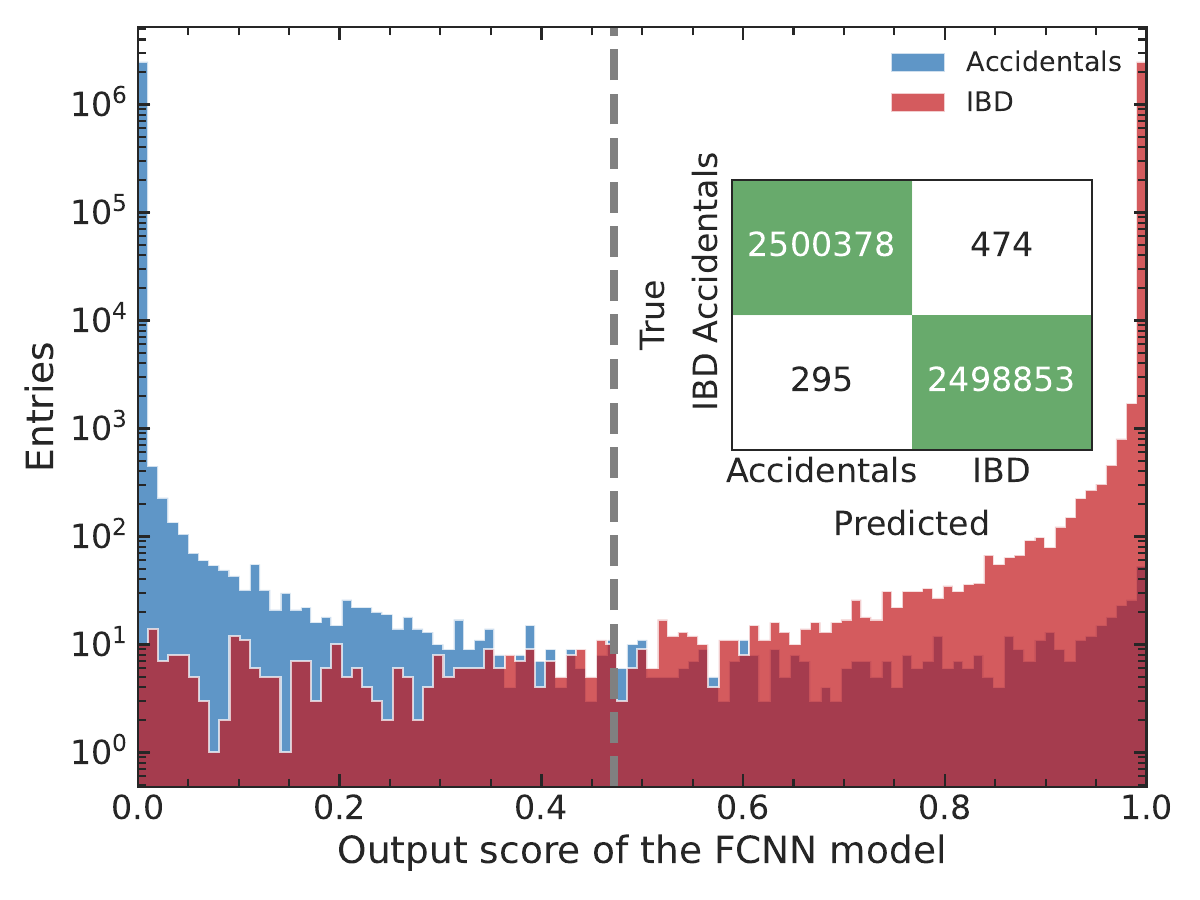}
    % \label{fig:fcnn_output}
    %(purity, efficiency, and F1-score)
	\caption{Left: Medians of the metrics by solid lines and their standard deviations after the bootstrap procedure as a function of the threshold (T-value). The best threshold value is shown with the dashed line. Right: %Probability to be an IBD events provided as output by the FCNN
	Output score provided by the FCNN model for events from the testing dataset. Most of the events are perfectly separated. The dashed line shows the best T-value. The inset plot represents a confusion matrix of the predictions.}
	\label{fig:fcnn_output_and_threshold}
\end{figure*}

\section{Results}
\label{sec:results}

One of the advantages of using a neural network {\color{arsenii}(as well as a BDT) to select IBD events is the non-binary output of the model. By applying the sigmoid function to the raw output, the models produce a value between 0 and 1 that can be associated with the model's confidence score of an event being an IBD candidate.
The models assign} this score to each event. The threshold $T$ in score above which an event is considered to be IBD is a tunable parameter. In absence of a prior physics requirement, one can choose a threshold based on balancing efficiency~(\autoref{eq:efficiency}) and purity~(\autoref{eq:purity}), maximizing the F1-score:
\begin{equation}
    \text{F1-score} = 2 \cdot \frac{\text{purity} \cdot \text{efficiency}}{\text{purity} + \text{efficiency}}
    \label{eq:F1}
\end{equation}
The F1-score is the harmonic mean of purity and efficiency and helps us strike a balance between correctly identifying accidental events (purity) and not missing any IBD events (efficiency). Given the small number of misclassifications, we use the bootstrap technique with the validation dataset to provide a more robust estimation of the optimal threshold value. This technique can be used to evaluate the variability of a parameter by repeatedly sampling from a dataset with replacement. In the context of our application, we re-sample 200 times the entire validation dataset (5M events) and evaluate purity, efficiency, and F1-score metrics at various threshold values. {\color{arsenii} As an example, we use a neural network (but closely the same result is obtained for the BDT) and to} assess the model's performance across different threshold values, we vary T from 0 to 1 in a uniform grid of 200 points. The left panel of~\autoref{fig:fcnn_output_and_threshold} reports the result of the evaluation procedure: the median values of purity, efficiency, and F1-score at different T-values are shown with the solid lines, while the corresponding standard deviation is represented by the shaded bandwidths. The best T-value, in terms of F1-score, is $\sim$0.47.~\autoref{fig:fcnn_output_and_threshold} depicts the FCNN results for the testing dataset and the dashed line stands for the chosen threshold value. As~\autoref{fig:fcnn_output_and_threshold} shows, by varying the $T$ value, one may vary signal to background ratio (and so the efficiency and purity of the selection). This may also be important in various physics analyses. For example, in physics channels where efficiency is the most significant metric, the $T$ threshold can be reduced to retain more IBD events, even if this results in degraded purity. On the other hand, where background hinders the estimation of the parameter of interest, it is important to balance the trade-off of these metrics. Using the optimized $T$ obtained from the maximization of the F1-score as the threshold to assign the IBD class, we get the following metrics for the model: efficiency of 99.988\%, purity of 99.981\%, and F1-score 99.985\%. This procedure for choosing a threshold serves as a generalized and agnostic method, and applicable with no specific requirement provided by a physics analysis. In the following subsection we instead introduce a physics-driven condition to fix a threshold $T$ for the model.

\subsection{ML selection and cut-based selection comparison}
\label{sec:ml_vs_standard}

{\autoref{tab:cuts_summary} presents a summary of the benchmark IBD selection applied to our dataset. It is worth noting that the outcome of this selection depends on the employed dataset, and may not reflect JUNO's official selection.}

\begin{table}[!h]
\renewcommand*{\arraystretch}{1.25}
    \centering
    \begin{tabular}{|c|c|}
    \hline
    Selection Criterion & Efficiency (\%)\\
    \hline
    All IBDs						& 100.0\\
    \hline
    FV cut 						& 91.7 \\
    \hline
    IBD Selection 				& 97.1 \\
    \hline
    ~~~~Energy cut 					& ~~~~98.7\\
    \hline
    ~~~~Time cut  		& ~~~~99.0\\
    \hline
    ~~~~Vertex cut 	& ~~~~99.4\\
    \hline
    Combined Selection					& 89.9\\
    \hline
    \end{tabular}
  \caption{Summary of the benchmark selection cuts and their single impact on IBD selection efficiency.}
    \label{tab:cuts_summary}
\end{table}

To evaluate the performances of the FCNN {\color{arsenii} and BDT models} and to compare it with the cut-based selection approach,  we use efficiency as the main metric with an additional condition on the background level, i.e., the fraction of selected (classified as IBDs) accidentals with respect to the total number. 

\begin{table*}[!htb]
\renewcommand*{\arraystretch}{1.25}
    \centering

    \begin{tabular}{|c|c|ccccc|} 
        \hline
        \multirow{2}{*}{\textit{Approach}} & \multirow{2}{*}{\textit{Volume}} & & & \textit{Efficiency}  & & \\\cline{3-7}
         &  & 0.2$\times$Bkg & 0.5$\times$Bkg & 1$\times$Bkg & 2$\times$Bkg & 5$\times$Bkg \\
        \hline
        \multirow{2}{*}{BDT} & Full detector volume: $R < 17.7$ m & 98.38\% & 98.81\% &\textbf{99.02}\% & 99.19\% & 99.39\% \\\cline{2-7}
                              &          $R < 17.2$ m & 91.58\% & 91.62\% & \textbf{91.63}\% & 91.64\% & 91.64\%  \\
        \hline
        \multirow{2}{*}{FCNN} & Full detector volume: $R < 17.7$ m & 96.94\% & 97.79\% &\textbf{98.40}\% & 98.82\% & 99.21\% \\\cline{2-7}
                              &          $R < 17.2$ m & 91.53\% & 91.60\% & \textbf{91.63}\% & 91.64\% & 91.64\%  \\
        \hline
        Cuts             &          $R < 17.2$ m & --- & --- & 89.90\% & --- & --- \\
        \hline
    \end{tabular}
    \caption{{\color{arsenii}The resulting efficiencies of the ML models and the comparison with the cut-based selection. Different background levels are used for comparison: equivalent (1$\times$Bkg), doubled (2$\times$Bkg), fivefold (5$\times$Bkg), halved (0.5$\times$Bkg), and five times lower (0.2$\times$Bkg). For the ML approaches, different background levels are adjusted to the corresponding ones by changing the threshold value. Moreover, we consider two cases for the neural network-based selection and BDT-based selection: (1) the FV cut applied, and (2) the full target volume. The muon veto cut is not included because it yields the same effect in all approaches.}}
    \label{tab:performances_table}
\end{table*}

Since only an extremely small number of accidentals satisfy the selection criteria, a very large radioactive sample is required to achieve a quantitative assessment of the two approaches. Therefore, an additional dataset was prepared, consisting of 147.5 million accidental coincidence pairs, corresponding to more than 1 year of data collection~\cite{bib:juno-radio}. Combined with the testing dataset, it consists of 152.5 million events, with 2.5 million events being IBD and the rest being accidentals. To ensure a comparison between {\color{arsenii} the ML models and the cut-based selection, the models are required} to achieve the same background level as the other approach, by choosing a specific threshold. Adhering to this requirement allows us to directly compare the efficiencies.

\autoref{tab:performances_table} shows the performances {\color{arsenii}of the models and their comparison with the cut-based selection.} Two different fiducial volumes are considered {\color{arsenii}for the ML models}: (1) the full target volume, $R < 17.7$ m, and (2) within the FV cut, i.e., $R < 17.2$ m. In the first case, thanks to greater flexibility and the ability to work with events at the detector edge, the ML models demonstrate a higher efficiency in tagging IBD events compared to the conventional approach, achieving an improvement of {\color{arsenii}$\sim$9.1 (BDT) and $\sim$8.5 (FCNN) percentage points.} At the same time, the background level is adjusted to the same value of $\sim$\num{1.27e-06} in all the approaches.
{\color{arsenii} Moreover, even within the FV volume, the boosted decision trees and the neural network are able to increase the efficiency of tagging IBD events both by $\sim$1.7 percentage points.} This increase comes from events which would be otherwise lost because of likely sub-optimal time and vertex cuts.

\begin{figure}[!t]
	\centering
	\includegraphics[width=0.9\columnwidth]{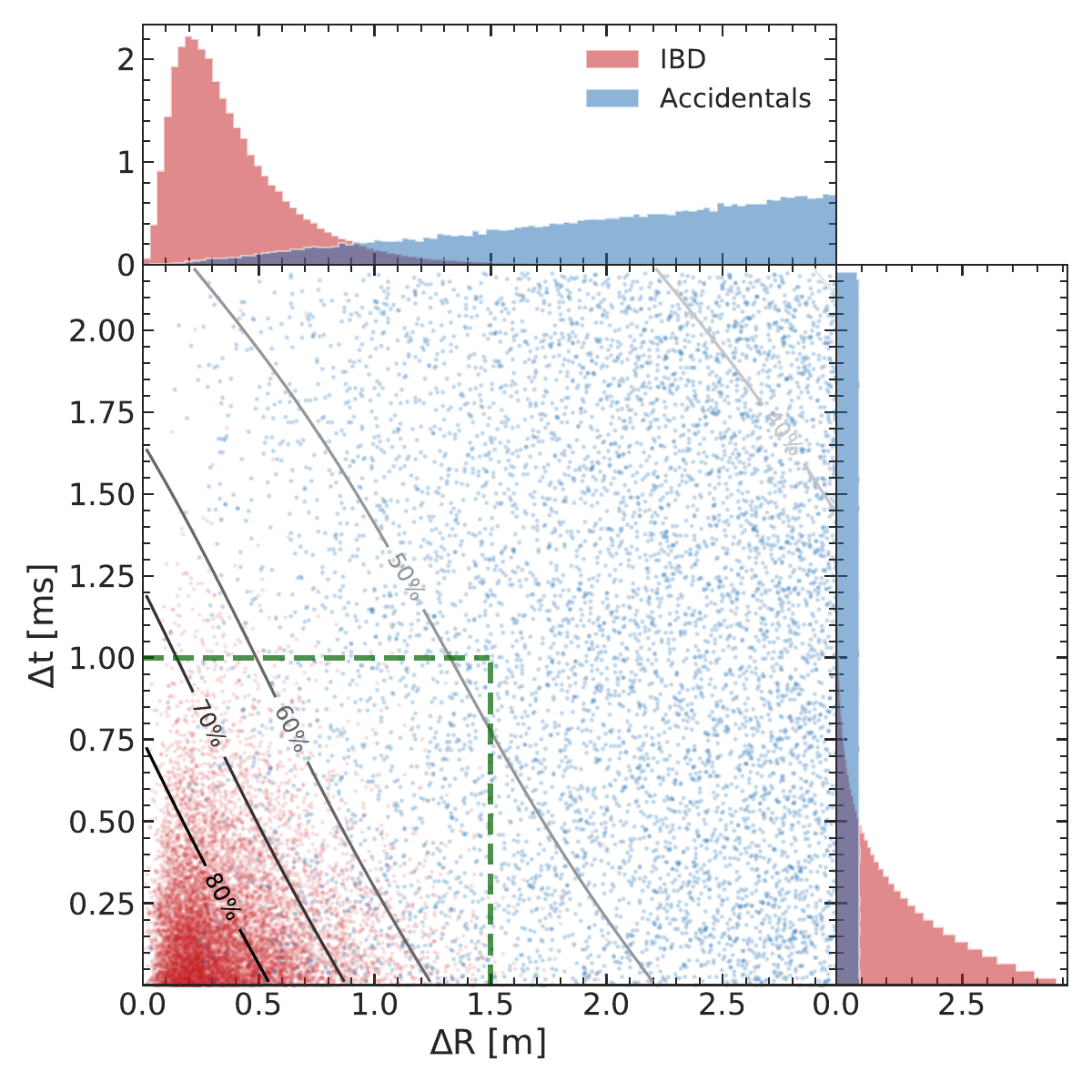}
	\caption{An example of a partial dependence plot for the following two features: $\Delta R$, $\Delta t$. Solid lines represent different FCNN's confidence levels that an event belongs to the \textit{IBD} class. The green lines show the cuts selection criteria. Blue and red points show the events from the testing dataset.}
	\label{fig:pdp}
\end{figure}

One approach to assess the influence of features on a model's output and their interconnections is the partial dependence plot
(PDP)~\cite{bib:pdp}. PDP computes the impact of a specific subset of features, typically one or two, by marginalizing (averaging) over all other features in a given feature set. Assuming $f$ represents a classifier, $\mathbf{x}_s$ denotes a set of feature values to be used in the evaluation of the PDP, $\mathbf{x}_c$ the remaining features, the partial dependence of $\mathbf{x}_s$ can be estimated as follows:
\begin{equation*}
    \hat{f}_s(\mathbf{x}_s) = \frac{1}{n}\sum_{i=1}^{n}f(\mathbf{x}_s, \mathbf{x}_{c, i}),
\end{equation*}
where $n$ is the number of events sampled from the training dataset, and $\mathbf{x}_{c, i}$ are values of the corresponding features.~\autoref{fig:pdp} shows an example of a partial dependence plot for the FCNN model, where $\mathbf{x}_s = (\Delta R, \Delta t)$.  
Here, we use the PDP to illustrate the interconnection between $\Delta R$ and $\Delta t$ and to compare the learned relationship with the cuts. {\color{arsenii} As was mentioned in Section~\ref{sec:problem_statement},} in contrast to the box-like decision boundary ($\Delta R < 1.5$ m, $\Delta t < 1$ ms), the neural network is able to learn a smoother boundary between the two classes, and therefore to improve the efficiency.\footnote{\color{arsenii} While BDT has learned an efficient separation boundary, its non-smooth, step-like nature makes it impractical for use as an optimized cut for the cut-based approach.}

\begin{figure}[t]
	\centering
	\includegraphics[width=1\columnwidth]{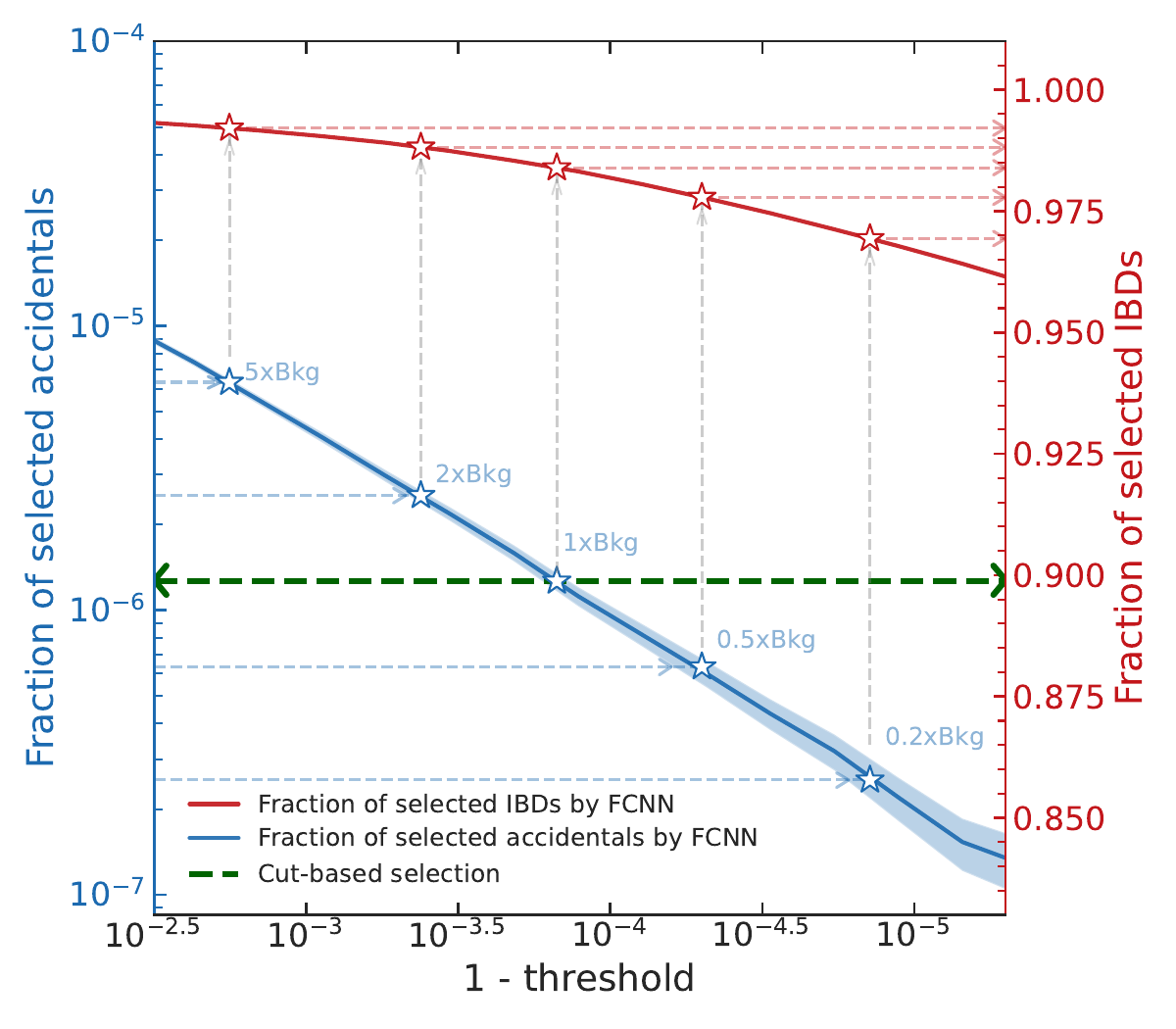}
	\caption{Comparison of the FCNN model performances and the cut-based selection for different threshold values. The green dashed line depicts the cut-based selection performances. The red and the blue solid lines show the fraction of the selected IBDs (efficiency) and the fraction of the selected accidentals (background level), respectively. Here, \textit{selected} is defined as \textit{classified as IBDs}. By relaxing the threshold, the model can achieve higher efficiency but obtaining more accidental events. The blue dashed lines point to the star markers that indicate the considered background levels, namely, equivalent (1$\times$Bkg), doubled (2$\times$Bkg), fivefold (5$\times$Bkg), halved (0.5$\times$Bkg), and five times lower (0.2$\times$Bkg). The red dashed lines show the corresponding efficiencies.}
	\label{fig:ml_vs_standard_selection}
\end{figure}

Furthermore, four additional scenarios are considered: a background level two times higher (2$\times$Bkg), five times higher (5$\times$Bkg), two times lower (0.5$\times$Bkg), and five times lower (0.2$\times$Bkg) than the one provided by the benchmark selection (1$\times$Bkg). As mentioned earlier, for certain physics analyses, a relatively elevated background level may not be critical, while the additional signal events are important and vise versa.
{\color{arsenii}\autoref{tab:performances_table} shows the change in efficiencies for these four scenarios: an increase of $\sim$0.17 percentage points (BDT) and $\sim$0.42 percentage points (FCNN) for the 2xBkg, and an increase of $\sim$0.37 percentage points (BDT) and $\sim$0.81 percentage points (FCNN) for the 5xBkg. Additionally, there is a decrease of only $\sim$0.21 percentage points (BDT) and $\sim$0.61 percentage points (FCNN) when suppressing the background level by two times, as well as a decrease of $\sim$0.64 percentage points (BDT) and $\sim$1.46 percentage points (FCNN) when suppressing the background level by five times.~\autoref{fig:ml_vs_standard_selection} depicts the dependence of the fraction of selected IBDs (efficiency) and the fraction of selected accidentals (background level) on different thresholds. The FCNN model is used as an example, with similar observations applicable to the BDT model. The green line shows the cut-based selection performances. The red dashed lines represent the efficiencies of the FCNN model with different background level conditions (illustrated by the blue dashed lines and star markers). Therefore, in the 1$\times$Bkg case, the difference between the green line and the red solid line indicates the increase in signal events statistics with respect to the cut-based selection.}

\section{Model's interpretability}
\label{sec:xai}

The \textit{black-box} nature of ML models can be overcome by employing constructs such as the Shapley values, introduced in the mid-20th century by Lloyd Shapley within the domain of cooperative game theory~\cite{bib:shap}. It stands as a measure to assess the importance of individual players within a coalition in reaching a common objective~\cite{bib:bookinterpretable}. 
Conceptually, the Shapley value gauges the impact of a player by quantifying how the average outcome changes when that player is included in the game, as opposed to its absence. It also serves as a fairness criterion, ensuring that each participant gains at least as much as they would have independently. Therefore, it is a valuable tool in situations where contributions are unequal, yet cooperation among players is essential to achieve a collective payoff~\cite{bib:bookinterpretable}. Mathematically, Shapley values provide a means to study the correlations between different variables. By considering all possible combinations of variables entering or leaving the game one can systematically evaluate their impact on the outcome. The main disadvantage lies in the fact that the exact calculation of Shapley values is challenging and requires extensive computation time~\cite{bib:bookinterpretable}.
In light of this challenge, for our study, we adopt the SHAP (SHapley Additive exPlanations) framework~\cite{bib:shap-implementation}.  SHAP~\cite{bib:shap-tree-nature, bib:shap-tree} introduces simplifications to address the computational challenges and speed up calculations while maintaining the interpretability and fairness of feature attributions~\cite{bib:shap-implementation,bib:shap-tree-nature, bib:shap-tree}. 
Explanations provided by SHAP offer valuable insights into the contribution of individual features to model outcomes, facilitating a comprehensive understanding from both global and local perspectives. Indeed, this framework enables us to explore not only the overall importance of features across the entire dataset but also the specific influence of features on individual predictions. 
{\color{arsenii}  Although SHAP values are a model-agnostic method that can be applied to both FCNN and BDT, we use FCNN for further analysis, as it is our main approach and provides all the desired features together.}
However, it is important to acknowledge that SHAP values may not perfectly capture the complexity of non-linear models, especially in deep learning models.

\subsection{Global and local explanations}

\begin{figure*}[!htb]
	\centering
	\begin{subfigure}{0.49\textwidth}
	\includegraphics[width=\textwidth]{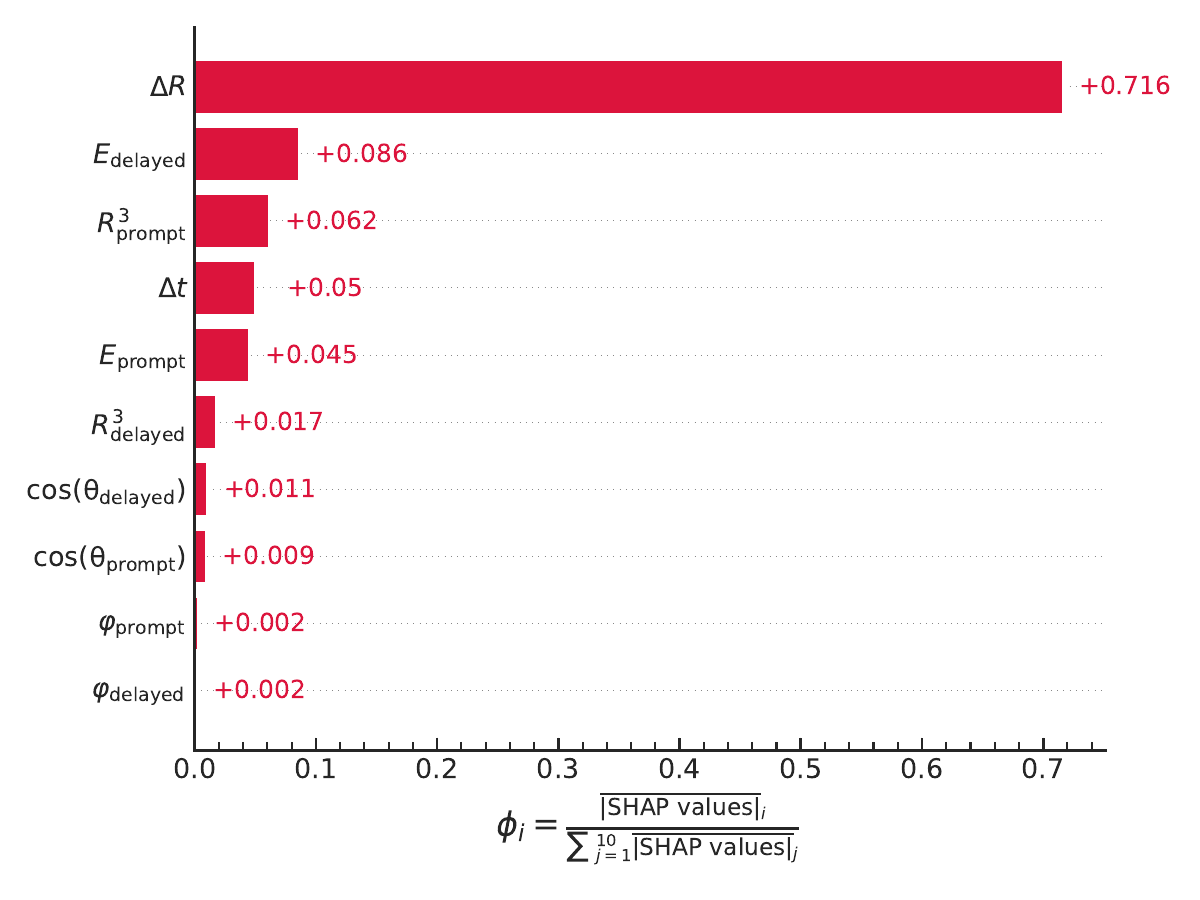}
%	\label{fig:glob_exp_shap}
    \end{subfigure}
	\hfill
	\begin{subfigure}{0.49\textwidth}
	\includegraphics[width=\textwidth]{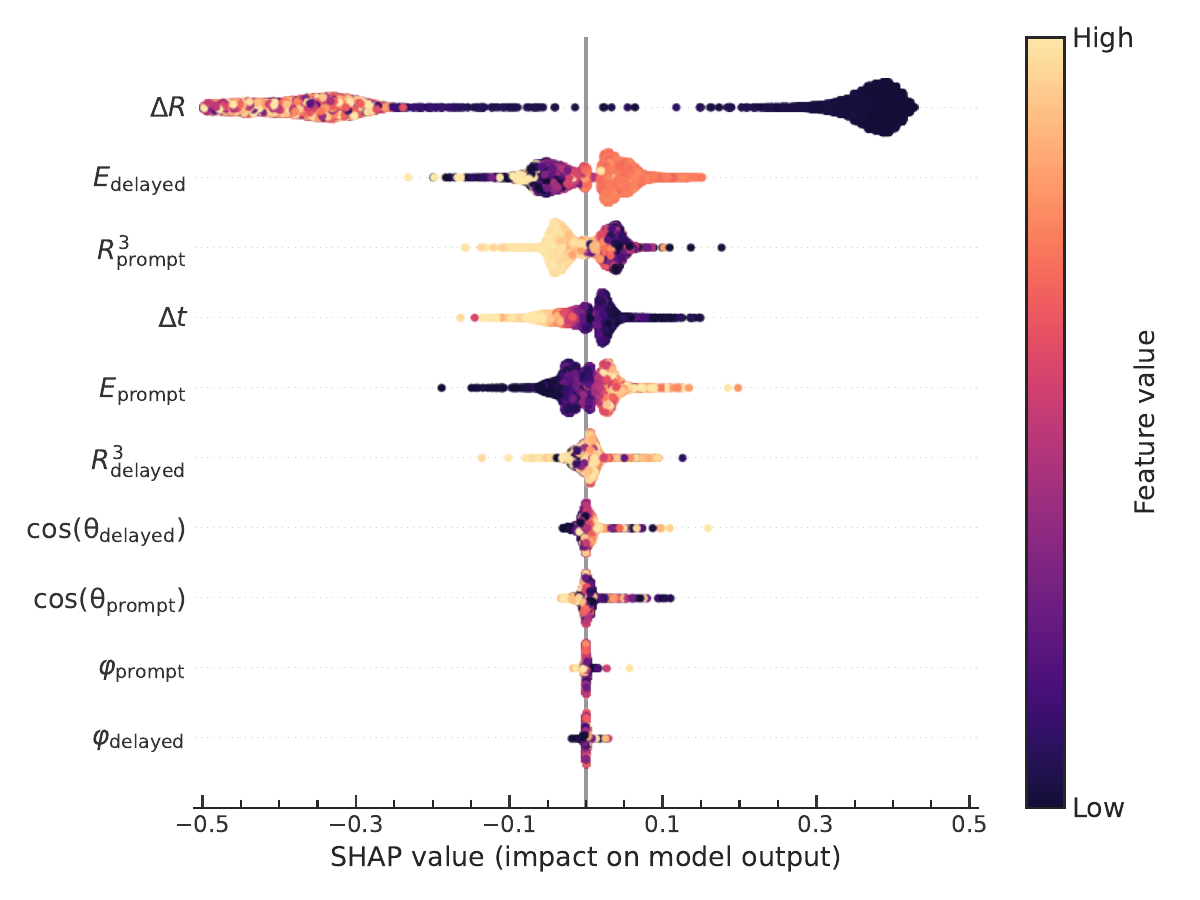}
%	\label{fig:loc_exp_shap}
    \end{subfigure}
	\caption{SHAP-based interpretability plots for the FCNN model. The left one represents the global explanation that summarizes the impact of the features on the model's output.  Notice that global explanations are normalized to $1$, i.e., $\sum_{i}^{10}{\phi_i} = 1$, where $i$ runs over the feature indexes. The right plot illustrates the local explanations for individual predictions and 20k events are shown. One event is represented by a point in each row, displaying the SHAP value associated with the corresponding feature. The density of SHAP values for a given feature represented by ``clumps''. The color provides the relative value of a feature: dark blue for low values and light blue for high values. Further details are provided in the text.}
	\label{fig:loc_glob_exp_shap}
\end{figure*}

\begin{figure*}[t]
	\centering
	\includegraphics[width=0.9\textwidth]{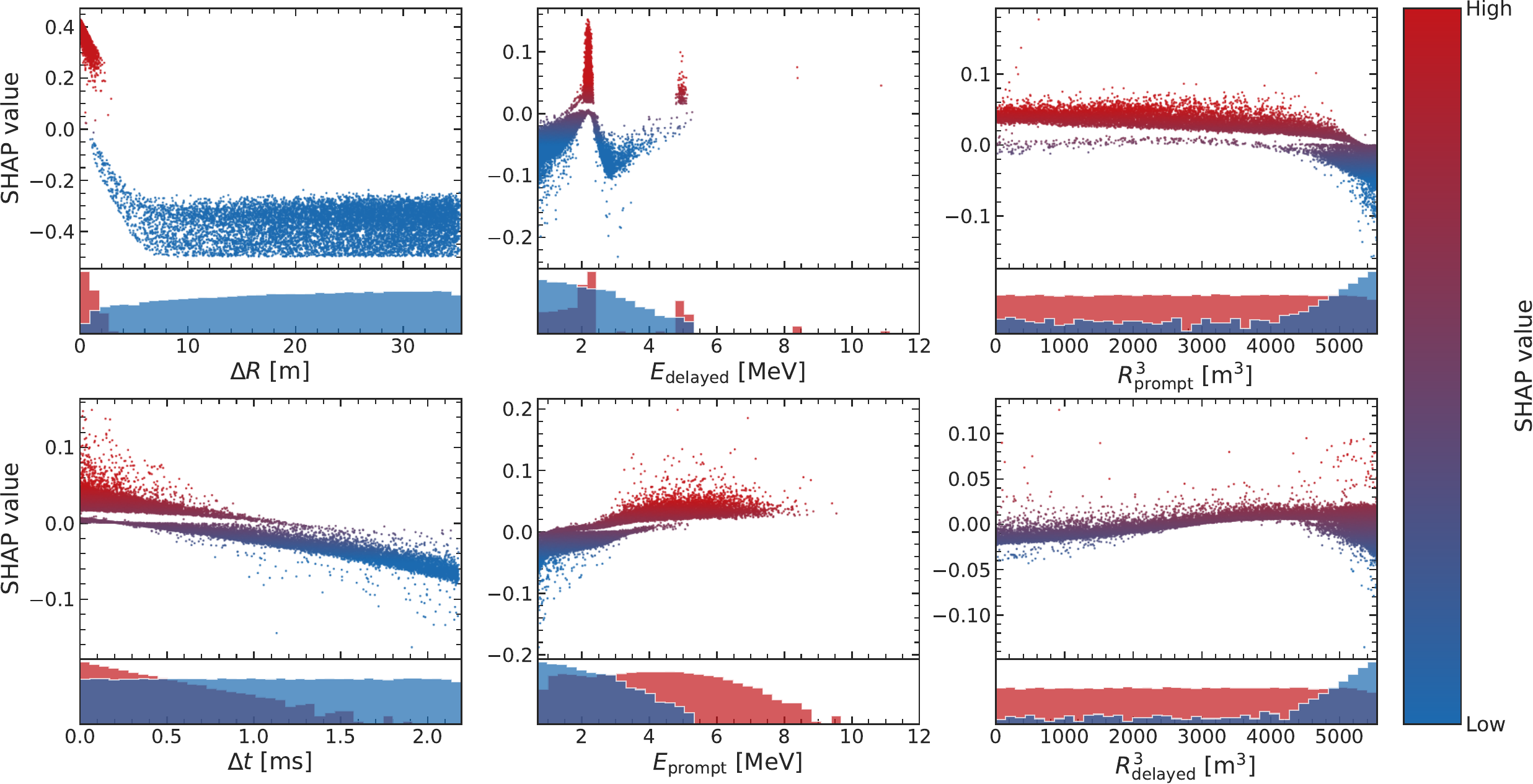}
	\caption{Detailed local explanations (top panel) for top-6 features and their distributions (bottom panel). Color represents SHAP value: from blue (negative, more confident to be accidentals) to red (positive, more confident to be IBD). The normalization of the color scale is set to enhance the differences between the two classes in terms of SHAP values.}
	\label{fig:shap_values_detailed}
\end{figure*}

\textit{Global} explanations compute the summarized impact of each input feature on the model output. Thanks to this, the generalized importance of the features can be estimated. It helps answer questions such as, ``What features are the most important for the model's predictions on average?''. Formally, global explanations are averaged absolute SHAP values calculated based on a provided data sample. On the other hand, \textit{local} explanations focus on the individual event and provide features' importance for a specific instance. It helps answer questions like, ``Why did the model make this particular prediction for this specific data point?''
We compute SHAP values for 20k events from the testing dataset. SHAP values can be both positive and negative, showing the impact on predictions with respect to the average value of the output variable (the labels 0 or 1 in a binary classification case). Since our datasets are balanced by construction, the mean value is equal to $0.5$. Thus, positive SHAP values indicate the contribution of a feature to pushing the model's output towards the IBD class. On the other hand, negative SHAP values indicate a contribution towards the \textit{accidentals} class.
 
The left part of~\autoref{fig:loc_glob_exp_shap} illustrates the global explanations for the FCNN model.\footnote{\color{arsenii} It is worth noting that another independent method for assessing feature importance ranking --- permutation importance~\cite{bib:perm_imp} --- provides nearly identical results.}
Each row corresponds to a feature and the bars' width represent the feature importance. The values are normalized to $1$, so that $\sum_{i}^{10}{\phi_i} = 1$. The most important features are, in order, $\Delta R$, $E_{\rm delayed}$, $R_{\rm prompt}^3$, and $\Delta t$. The impact of $E_{\rm prompt}$ and $R_{\rm delayed}^3$, being smaller on average, helps in the selection of rarer cases. The same applies to the cosine theta features, which allow the model to correct the prediction, especially for values at their extremes. In contrast to $\cos(\theta)$, the azimuthal angles $\varphi_{\rm prompt}, \varphi_{\rm delayed}$ features have almost negligible importance. {\color{arsenii} We would also like to highlight that the independent method for assessing global feature importances --- gain feature importance for BDT --- is mostly in agreement with the results obtained with SHAP with FCNN. Both methods identify the same three main clusters of features: 
 $\Delta R$, radii, energies and $\Delta t$, followed by the angles-related features.}

Local explanations help to better understand why certain features are more or less important and in which cases. The right panel of~\autoref{fig:loc_glob_exp_shap} illustrates a set of SHAP values for each feature for 20k events taken from the testing dataset. One event is a point in each row, hence it is decomposed into ten points. The color represents the relative value of a feature: from dark blue (values are close to its minimum) to light blue (values are close to its maximum). The concentration of events on certain SHAP values is shown as ``clumps''. Regarding $\Delta R$, there is a clear correlation between its value, (i.e., the color of a data point), and the corresponding SHAP value (i.e., the position on the horizontal axis): for events with smaller $\Delta R$ the model is more confident to assign the \textit{IBD} class (positive SHAP values) than for events with large $\Delta R$ (negative SHAP values).
The next most important feature is $E_{\rm delayed}$ because its distribution is different for the two classes: in particular, the clustered structure (related to the different gamma emission peaks) that we observe for IBD events is not present for accidentals. For IBD events, it is strictly related to the isotope that captures the IBD neutron: 2.2 MeV ($^1$H), 4.95 MeV ($^{12}$C), higher energies ($^{13}$C, $^{14}$N). Thus, positive SHAP values are associated with events with these particular $E_{\rm delayed}$. While the cut-based selection completely reject events with higher $E_{\rm delayed}$ energies, the FCNN model is able to preserve them, increasing efficiency. Another energy-related feature, $E_{\rm prompt}$, has the following dependence: at small values, the model is more confident that the events belong to the \textit{accidentals} class since this part of the energy spectrum is populated mainly by the $^{14}$C isotope, having very few events associated with reactor antineutrinos. On the contrary, accidentals with higher energies are almost nonexistent and IBDs dominate, resulting in positive SHAP values.

\autoref{fig:shap_values_detailed} shows detailed explanations for the top-6 features: for each event, the top panel reports the feature value on the $x$-axis, with the corresponding SHAP value on the $y$-axis. The bottom panel shows histograms of the feature distributions, for accidental coincidences in blue and IBD pairs in red, as previously reported in \autoref{fig:feature_distributions}. The colorbar shows the relative contribution for these features in terms of SHAP value, while also being a proxy for the y-axis values. 
These explanations are useful to visualize the relation between the feature distributions and their impact on the model's output. Bright red and blue regions correspond to features significantly pushing the model to the \textit{IBD} and \textit{accidentals} classes, respectively. On the other hand, purple-shaded areas provide little contribution to the model's output.
For example, as was mentioned above, $E_{\rm delayed}$ explanations have a clear clustered structure. There are several clusters of positive SHAP values associated with the released gamma energies from neutron capture on different isotopes. The width of the cluster can be used as cut boundaries for the benchmark selection procedure.
~\autoref{fig:shap_values_detailed} also shows a clear clustering structure for $\Delta R$ and $E_{\rm prompt}$. Indeed, for events with $\Delta R \lesssim$~2~m the model is more confident to assign the \textit{IBD} class, while for events with $\Delta R \gtrsim$~2~m, FCNN is less confident, resulting in negative SHAP values.
Concerning $E_{\rm prompt}$, SHAP values follow the feature distributions and have an overlapping in the region of [1.5, 4]~MeV. In this energy range, correlations with other features play a key role (mainly $E_{\rm delayed}$ and $\Delta R$), allowing the model to distinguish IBDs from accidentals. 
The time-related feature $\Delta t$ mostly pushes towards the \textit{IBD} class when it is within several neutron capture times $\tau$. On the other hand, in the case of highly separated in time events, they are considered more probable to be accidentals. Concerning the position-related features ($R_{\rm prompt}^3$, $R_{\rm delayed}^3$) in \autoref{fig:shap_values_detailed}, their absolute values become more significant closer to the edge, increasing the confidence that an event is accidental.
It is interesting to note that some events have a large positive SHAP value for $R^3_{\rm delayed}>17.65^3$ m$^3$. This is because, in the case of events at the very edge of the detector, the probability of gamma leakage becomes higher. In a liquid scintillator detector, gamma leakage refers to the energy loss caused by gamma rays not depositing all their energy within the detector volume. Even though the energy of the delayed event is much less than expected, the event is correctly classified as IBD according to the values of all other features, including information about proximity to the edge.

\subsection{Special cases}

Local explanations, which provide insights into how the model makes decisions for individual events, are an effective tool for debugging the model and identifying special cases.  In order to do this, we employ the so-called \textit{waterfall} plot~\cite{bib:shap-implementation}, a visualization tool that conveys the impact of SHAP values on our model's output. The $x$-axis reports the expected value of the model output $E[f(x)]$: it starts from a baseline value, set at 0.5 in our case due to dataset balance\footnote{If we randomly sample an event from the dataset, we have a 50\% probability for it to be either an IBD or an accidental coincidence.}, and each subsequent row shows how each feature contributes to the overall prediction. The color indicates whether a specific feature pushes the prediction higher (red, i.e., more confident to be IBD) or lower (blue, thus more confident to be background) than the base value.

\begin{figure*}[t]
	\centering
	\begin{subfigure}{\textwidth}
	\includegraphics[width=0.49\textwidth]{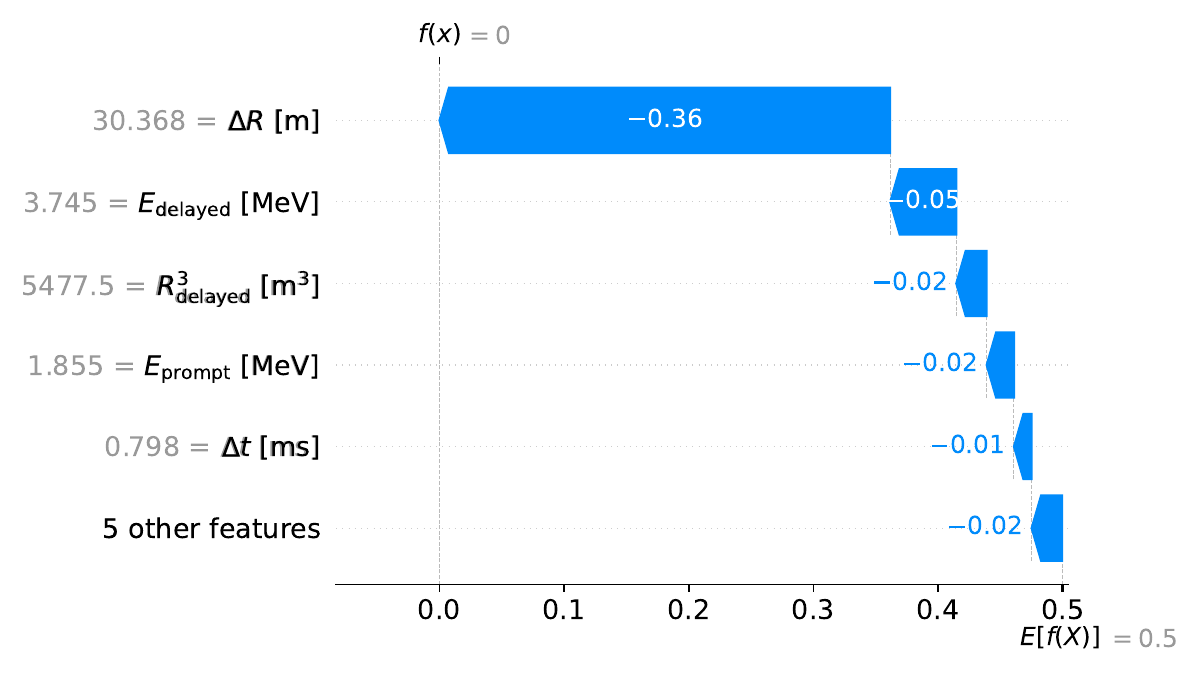}
	\includegraphics[width=0.49\textwidth]{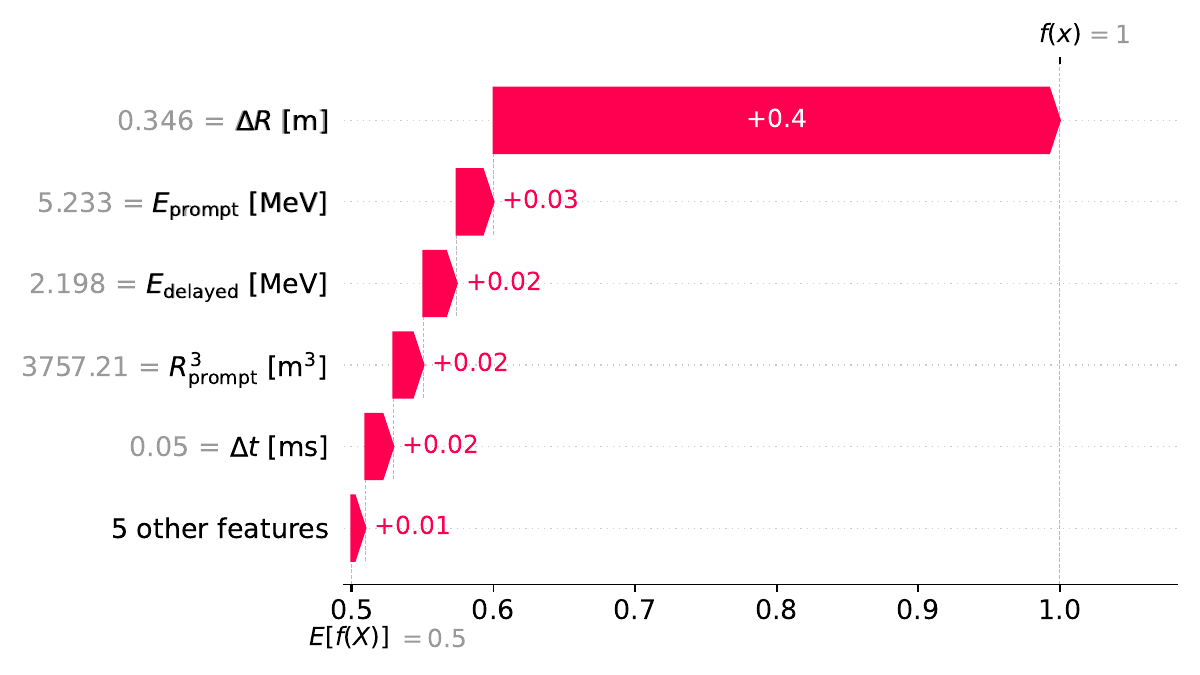}
	\caption{SHAP interpretability plot of FCNN predictions for typical cases of accidentals (left) and IBD (right). }
	\label{fig:shap_cases_typical}
    \end{subfigure}
	\hfill
	\begin{subfigure}{\textwidth}
	\includegraphics[width=0.49\textwidth]{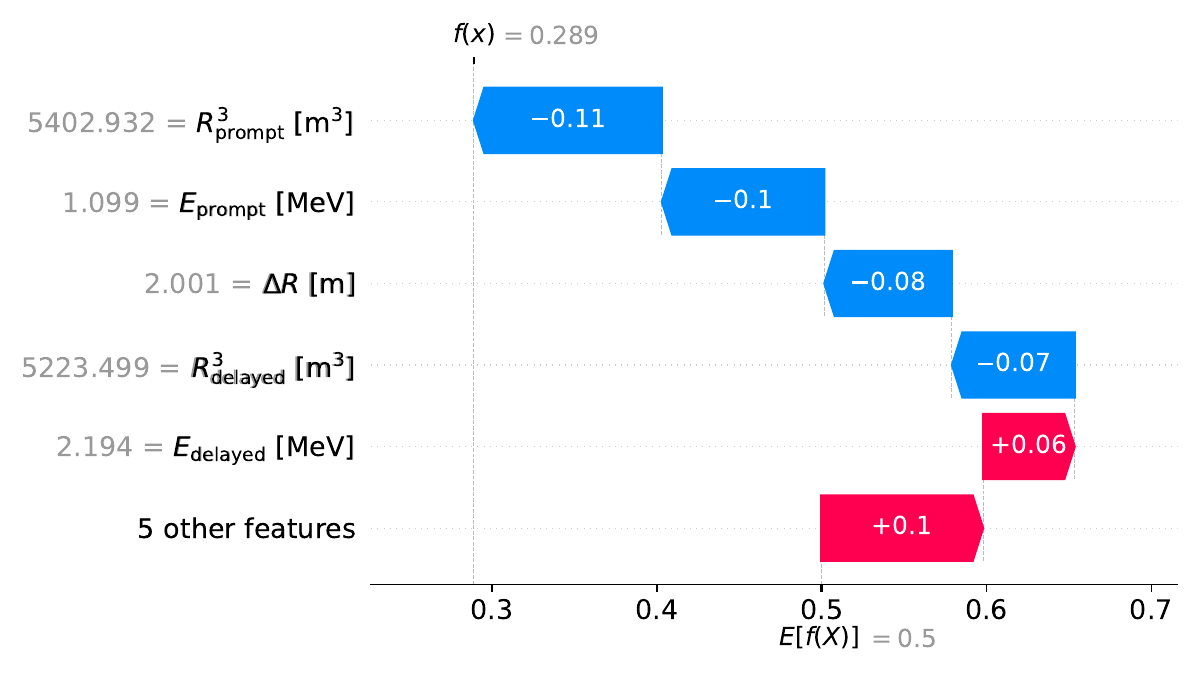}
	\includegraphics[width=0.49\textwidth]{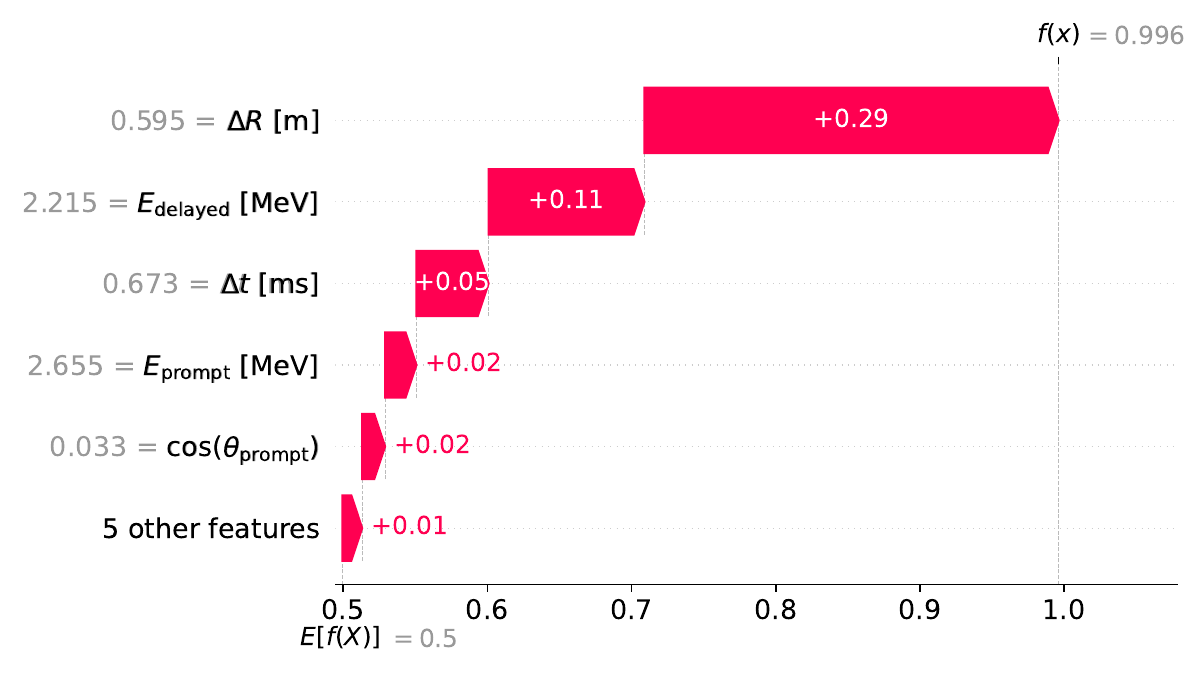}
	\caption{SHAP interpretability plot for cases of wrong predictions of FCNN for true IBD (left) and true accidentals (right).}
	\label{fig:shap_cases_wrong}
    \end{subfigure}
	\caption{SHAP explanations provided for particular cases of correct classifications (a) and misclassifications (b). Features are sorted based on the magnitude of their SHAP values, and the smallest magnitude features are clustered at the bottom of the plot.}
	\label{fig:shap_cases}
\end{figure*}

Focusing on specific cases, ~\autoref{fig:shap_cases_typical} reports SHAP values for events that were correctly classified, namely a typical accidental coincidence on the left panel and an IBD pair on the right side.
The accidental event has a large $\Delta R$ of $\sim$30~m and $E_{\rm delayed}$ outside the energy ranges of neutron capture gammas. This combination of values already allows the model to designate this event as a random coincidence with $\sim$90\% confidence. On the other hand, for the right event with $\Delta R = 0.35$~m, $E_{\rm {prompt}} = 5.23$~MeV, and $E_{\rm {delayed}}$ $\sim$2.2~MeV the model assigned a $\sim$100\% confidence score to be an IBD event.

\begin{figure}[!htb]
	\centering
	\includegraphics[width=1\columnwidth]{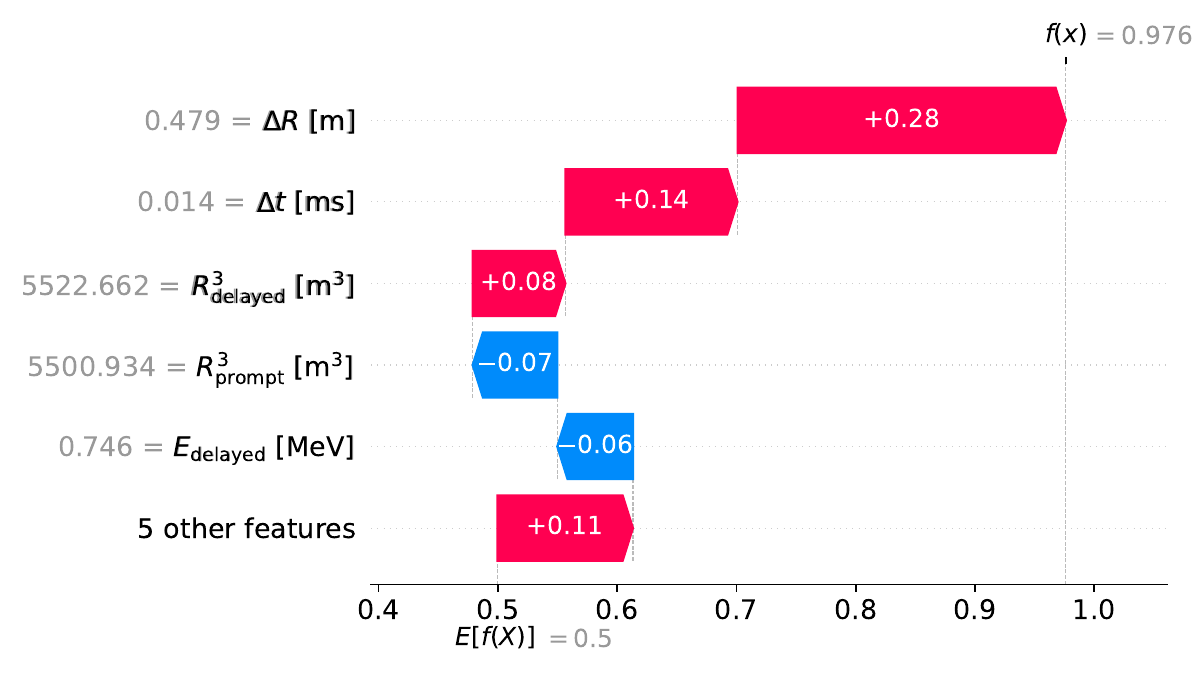}
	\caption{An example of an IBD event with the gamma leakage effect: the gamma produced by neutron capture did not deposit its entire energy in the LS, but instead escaped the target. Despite the almost complete energy leakage ($E_{\rm delayed}$ of 0.746 MeV), the model classifies this as an IBD event with $\sim$100\% confidence, based on the combination of other features.}
	%probability
	\label{fig:gamma_leakage}
\end{figure}

Moreover,~\autoref{fig:gamma_leakage} shows an example of a correctly classified gamma leakage event that would be discarded by the cut-based selection because of the very low $E_{\mathrm{delayed}}$ and the FV cut. Contrariwise, the ML model is able to identify this kind of events thanks to the combination of all other features. The efficiency of classification of events with the gamma leakage effect is $\gtrsim$~95\%, using the threshold optimized based on F1-score maximization.

On the other hand,~\autoref{fig:shap_cases_wrong} illustrates cases when the model made a wrong prediction. The left one is a true IBD with an escaped gamma that was classified as an accidental event. Unlike the gamma leakage event presented above in~\autoref{fig:gamma_leakage}, this event has a low classification score mostly because of the atypical high $\Delta R$. The right panel of~\autoref{fig:shap_cases_wrong} shows a true accidental event that was classified as IBD. This misclassification was caused by the unlikely case of accidentals with an IBD-like combination of features.

\section{Model calibration and uncertainty quantification}
\label{sec:mc_ue}
{\color{arsenii} In this section, we discuss two important aspects of building a trustworthy and reliable machine learning model: model calibration and uncertainty estimation.

Model calibration refers to the process of aligning the output scores predicted by a classifier to probabilities~\cite{bib:mc1}.}
{\color{andrea}A well-calibrated model produces scores that can be directly interpreted as probabilities, allowing for straightforward use of the classifier's predictions in situations where a probabilistic interpretation is a desired and pivotal output. In these cases, the relationship between probability and classifier output score appears as a diagonal straight line with unit slope and zero intercept. Uncalibrated models may deviate from this straight line, associating high output score values with low probabilities (i.e., overconfident classifiers), or conversely associating low score values with high probabilities (i.e., underconfident), or combinations of the two. Model calibration is not always a fundamental requirement, but it greatly simplifies the interpretation of model decisions. Finally, calibration is not an intrinsic property of the model, but also depends on the reference dataset. Various techniques exist to calibrate uncalibrated models, should this be required during use~\cite{bib:calibration_of_mnn}.}

{\color{andrea}The second aspect to address is uncertainty estimation.}
{\color{arsenii} In the machine learning literature, uncertainty estimation is often divided into two types: (i) aleatoric and (ii) epistemic uncertainty~\cite{bib:ml_unc_review}. Aleatoric uncertainty is related to the inherent randomness in the data itself, making it irreducible~\cite{bib:ml_unc_review}. In contrast, epistemic uncertainty refers to the intrinsic model uncertainty, which is potentially reducible by providing additional information during the training of the model~\cite{bib:ml_unc_review}. Understanding and quantifying uncertainty not only improves model interpretability but also provides a measure of confidence in its predictions. This is particularly important when dealing with areas of sparse data or when identifying out-of-distribution events, where predictions are likely to be less reliable. Several methods exist for quantifying the impact of model uncertainty~\cite{bib:ml_unc_methods}. An effective and relatively simple to implement method for estimating epistemic uncertainty is Monte Carlo (MC) dropout~\cite{bib:mc_dropout}.

MC dropout is a technique that uses dropout~\cite{bib:dropout} to approximate Bayesian inference during both model training and inference. In standard dropout, neurons of the network are randomly ''dropped'' (i.e., set to zero) during training with a probability $p$ for each neuron, and then at inference stage, dropout is turned off. However, in MC dropout, dropout is kept active during inference as well: by performing multiple forward passes, it enables the model to make predictions under slightly different conditions (different set of active neurons) for each pass, allowing us to capture the variance in these predictions as a measure of uncertainty. Thus, MC dropout keeps the stochasticity at inference stage as well.

In this study, we applied dropout layers to our FCNN classifier after each fully connected layer, using a dropout probability of $p = 0.2$. For each event in our testing dataset, we performed 100 forward passes, recording the average prediction and its standard deviation. \autoref{fig:unc_est_calib_plot} presents a calibration plot showing both the epistemic uncertainty estimated using the MC dropout technique and the model’s calibration. On the x-axis, we display the fraction of IBD events (frequentist probability) in each bin and its statistical uncertainty, while the y-axis shows the corresponding averaged predictions and their standard deviations, representing epistemic uncertainty as estimated by MC dropout. This plot demonstrates that our model is well-calibrated across the testing dataset, as the predictions closely align with the diagonal line representing perfect calibration. Additionally, the uncertainty (represented by the error bars) is minimal at the edges (where predictions are closer to 0 or 1), and increases in the middle of the range. This increase in uncertainty around predictions near 0.5 can be attributed to the sparser data points in this region, which makes the model less confident in its predictions.

\begin{figure}[!htb]
	\centering
        \includegraphics[width=0.95\columnwidth]{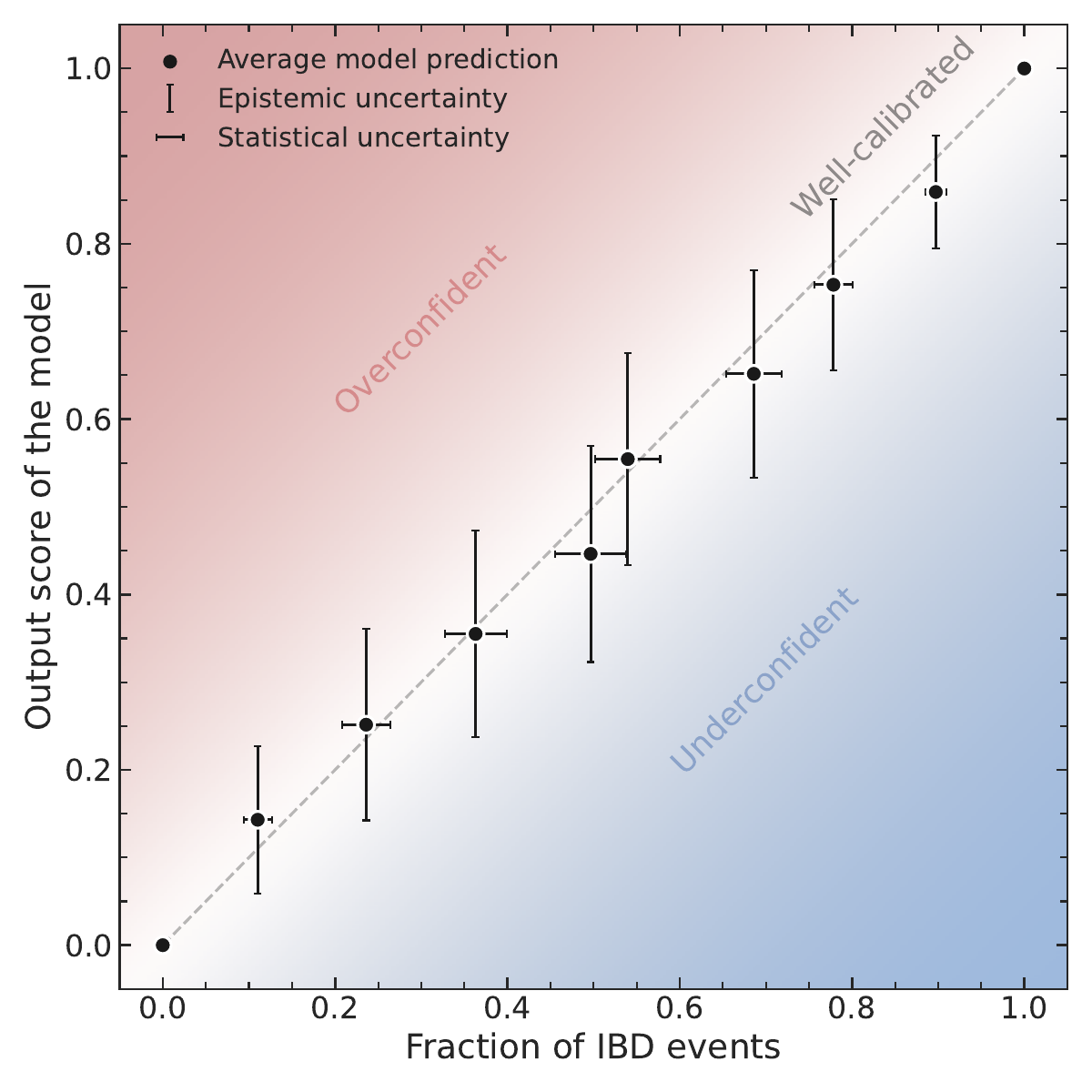}
	\caption{{\color{arsenii} Calibration plot for the FCNN model with the Monte Carlo dropout technique applied. The model is well-calibrated across the 5M events testing dataset. On the x-axis the fraction of IBD events in each bin (frequentist probability) and its statistical uncertainty are presented. The y-axis displays the average predictions and corresponding epistemic uncertainty provided by the MC dropout. The uncertainty estimation is done using 100 forward passes.}}
	\label{fig:unc_est_calib_plot}
\end{figure}

Additionally, in the case of a realistic class imbalance (with significantly more accidentals than IBD events), the same model will show overconfidence: a smaller fraction of IBD events will correspond to the same output score. However, this discrepancy can be easily resolved with a simple scaling procedure using the following expression: $p = \frac{s}{1 + (R - 1)(1 - s)}$, where $R$ is the class imbalance ratio: $R = \frac{N_{\rm accidentals}}{N_{\rm IBD}}$, $s$ represents the confidence scores provided by the model, $p$ is the resulting probability. We would like to note that the rates can be estimated independently from data before running the ML selection.}

\section{Conclusions}
\label{sec:conclusions}
In this study, we introduced a machine learning model, specifically a fully connected neural network, for event selection in a large liquid scintillator detector. Taking the JUNO experiment as a case study, we demonstrate that the presented ML model {\color{andrea} (i) is capable of matching the performance of a boosted decision trees based classifier} and (ii) is able to learn a more flexible boundary between signal and background events compared to a cut-based selection criteria.
This consequently leads to a $\sim$1.7 percentage points increase of efficiency within the fiducial volume.
Moreover, the ML approach opens up the possibility to remove the strict fiducial volume cut, retaining a higher number of signal events, providing an improvement of $\sim$8.5 percentage points in efficiency. For both cases, the model keeps exactly the same background level as the cut-based selection. It also proves to be powerful in tagging events characterized by gamma leakage, that would otherwise be discarded by the cuts.
Furthermore, we outline a systematic approach for preparing datasets and optimizing model hyperparameters. This methodology is not exclusive to JUNO but can be extended and applied to other liquid scintillator-based detectors, for any supervised learning tasks. 
A key aspect of our study involves interpretability analysis, aimed at investigating the decision-making process of the ML model and offering valuable insights into its behavior. This deepened understanding contributes to refining cut-based event selection strategies, ensuring the robustness of model predictions, both at the local (individual event) and global (across a set of events) levels. {\color{andrea} Part of the work has been devoted to model calibration and estimating the epistemic uncertainty of its predictions in order to corroborate the confidence in the network's decisions and outputs.}
In summary, our work underscores the potential of the ML approach to optimize event selection for inverse beta decay interactions in neutrino experiments. This flexibility proves particularly advantageous in striking a balance between purity and efficiency tailored to the physics channel of interest.

\section*{Acknowledgements}
We are thankful to the JUNO reactor antineutrino physics working group for the support and advice provided during the drafting of this manuscript. We are also grateful to CloudVeneto for providing IT support and GPU resources, and to the IHEP computing center for providing access to their computing cluster.
This project has received funding from the European Union’s Horizon 2020 research and innovation programme under the Marie Skłodowska-Curie Grant Agreement No. 101034319 and from the European Union – NextGenerationEU.

\bibliographystyle{JHEP} % 

\bibliography{bibliography.bib}

\providecommand{\href}[2]{#2}\begingroup\raggedright\begin{thebibliography}{10}

\bibitem{bib:ml-particlephysics}
D.~Bourilkov, {``Machine and deep learning applications in particle physics''}, \href{https://doi.org/10.1142/S0217751X19300199}{\emph{International Journal of Modern Physics A} {\bfseries 34} (2019) 1930019}.

\bibitem{ml-schwartz}
M.~D. Schwartz, {``{Modern Machine Learning and Particle Physics}''}, {\emph{Harvard Data Science Review} (2021) }.

\bibitem{bib:nova-pid}
D.~Rocco, M.~Messier, E.~Niner, G.~Pawloski and P.~Vahle, {``A convolutional neural network neutrino event classifier''}, \href{https://doi.org/10.1088/1748-0221/11/09/P09001}{\emph{Journal of Instrumentation} {\bfseries 11} (2016) P09001}.

\bibitem{bib:ml-next}
J.~Renner et~al., {``Background rejection in {NEXT} using deep neural networks''}, \href{https://doi.org/10.1088/1748-0221/12/01/T01004}{\emph{Journal of Instrumentation} {\bfseries 12} (2017) T01004}.

\bibitem{bib:juno-rec}
Z.~Qian et~al., {``{Vertex and energy reconstruction in JUNO with machine learning methods}''}, \href{https://doi.org/10.1016/j.nima.2021.165527}{\emph{Nucl. Instrum. Meth. A} {\bfseries 1010} (2021) 165527}.

\bibitem{bib:juno-rec_en}
A.~Gavrikov, Y.~Malyshkin and F.~Ratnikov, {``{Energy reconstruction for large liquid scintillator detectors with machine learning techniques: aggregated features approach}''}, \href{https://doi.org/10.1140/epjc/s10052-022-11004-6}{\emph{Eur. Phys. J. C} {\bfseries 82} (2022) 1021}.

\bibitem{bib:ratnikov_gans}
V.~Chekalina et~al., {``Generative models for fast calorimeter simulation: the {LHCb} case''}, \href{https://doi.org/10.1051/epjconf/201921402034}{\emph{EPJ Web Conf.} {\bfseries 214} (2019) 02034}.

\bibitem{bib:ratnikov_detectors}
{\scshape MODE} collaboration, {``{Toward the end-to-end optimization of particle physics instruments with differentiable programming}''}, \href{https://doi.org/10.1016/j.revip.2023.100085}{\emph{Rev. Phys.} {\bfseries 10} (2023) 100085}.

\bibitem{bib:mb_example}
{\scshape MicroBooNE} collaboration, \emph{{Reconstruction and Selection of Neutrino Interactions in MicroBooNE using Deep Convolutional Neural Networks}}. July, 2024, \href{https://doi.org/10.2172/2406059}{10.2172/2406059}.

\bibitem{bib:sk_example}
P.~Fernandez~Menendez, {``{Atmospheric neutrino oscillations with Super-Kamiokande and prospects for SuperK-Gd}''}, \href{https://doi.org/10.22323/1.395.0008}{\emph{PoS} {\bfseries ICRC2021} (2021) 008}.

\bibitem{bib:km_example}
T.~Sakai, \emph{{Advanced new tool for background rejection in KamLAND geo-neutrino analysis using machine learning methods}}. Neutrino24, June, 2024, \href{https://doi.org/10.5281/zenodo.13122055}{10.5281/zenodo.13122055}.

\bibitem{bib:ml-explainable}
V.~Belle and I.~Papantonis, {``Principles and practice of explainable machine learning''}, \href{https://doi.org/10.3389/fdata.2021.688969}{\emph{Frontiers in big Data} (2021) 39}.

\bibitem{bib:bookinterpretable}
C.~Molnar, \emph{Interpretable Machine Learning: A Guide for Making Black Box Models Explainable}. Online, 2022.

\bibitem{bib:majorana}
I.~Arnquist et~al., {``{Interpretable boosted-decision-tree analysis for the Majorana Demonstrator}''}, \href{https://doi.org/PhysRevC.107.014321}{\emph{Physical Review C} {\bfseries 107} (2023) 014321}.

\bibitem{bib:reines-cowan}
C.~L. Cowan~Jr et~al., {``Detection of the free neutrino: a confirmation''}, \href{https://doi.org/10.1126/science.124.3212.10}{\emph{Science} {\bfseries 124} (1956) 103}.

\bibitem{bib:vogel-xs}
P.~Vogel and J.~Beacom, {``Angular distribution of neutron inverse beta decay, $\bar{\nu}_e + p \rightarrow e^+ + n $''}, \href{https://doi.org/10.1103/PhysRevD.60.053003}{\emph{{Physical Review D}} {\bfseries 60} (1999) 053003}.

\bibitem{bib:kamland}
K.~Eguchi et~al., {``First results from {KamLAND}: evidence for reactor antineutrino disappearance''}, \href{https://doi.org/10.1103/PhysRevLett.90.021802}{\emph{{Physical Review Letters}} {\bfseries 90} (2003) 021802}.

\bibitem{bib:dayabay}
F.~An et~al., {``Observation of electron-antineutrino disappearance at {Daya Bay}''}, \href{https://doi.org/10.1103/PhysRevLett.108.171803}{\emph{{Physical Review Letters}} {\bfseries 108} (2012) 171803}.

\bibitem{bib:doublechooz}
Y.~Abe et~al., {``Indication of reactor $\nu_e$ disappearance in the {Double Chooz} experiment''}, \href{https://doi.org/10.1103/PhysRevLett.108.131801}{\emph{{Physical Review Letters}} {\bfseries 108} (2012) 131801}.

\bibitem{bib:RENO}
J.~K. Ahn et~al., {``Observation of reactor electron antineutrinos disappearance in the {RENO} experiment''}, \href{https://doi.org/10.1103/PhysRevLett.108.191802}{\emph{{Physical Review Letters}} {\bfseries 108} (2012) 191802}.

\bibitem{bib:junophysics2016}
F.~An et~al., {``Neutrino physics with {JUNO}''}, \href{https://doi.org/10.1088/0954-3899/43/3/030401}{\emph{Journal of Physics G: Nuclear and Particle Physics} {\bfseries 43} (2016) 030401}.

\bibitem{bib:junophysics}
A.~Abusleme et~al., {``{JUNO} physics and detector''}, \href{https://doi.org/10.1016/j.ppnp.2021.103927}{\emph{Progress in Particle and Nuclear Physics} {\bfseries 123} (2021) 103927}.

\bibitem{bib:nnn2023}
M.~Grassi, \emph{{The JUNO Experiment: Status and Prospects}}. NNN23, Oct., 2023, \href{https://doi.org/10.5281/zenodo.10624758}{10.5281/zenodo.10624758}.

\bibitem{bib:juno-radio}
A.~Abusleme et~al., {``Radioactivity control strategy for the {JUNO} detector''}, \href{https://doi.org/10.1007/JHEP11(2021)102}{\emph{Journal of High Energy Physics} {\bfseries 2021} (2021) 1}.

\bibitem{bib:sub_osc}
A.~Abusleme et~al., {``Sub-percent precision measurement of neutrino oscillation parameters with {JUNO}''}, \href{https://doi.org/10.1088/1674-1137/ac8bc9}{\emph{Chinese Physics C} {\bfseries 46} (2022) 123001}.

\bibitem{bib:junosw}
T.~Lin et~al., {``{Simulation software of the JUNO experiment}''}, \href{https://doi.org/10.1140/epjc/s10052-023-11514-x}{\emph{Eur. Phys. J. C} {\bfseries 83} (2023) 382}.

\bibitem{bib:bdt1}
J.~H. Friedman, {``{Stochastic gradient boosting}''}, \href{https://doi.org/10.1016/S0167-9473(01)00065-2}{\emph{Comput. Stat. Data Anal.} {\bfseries 38} (2002) 367}.

\bibitem{bib:bdt2}
J.~H. Friedman, {``{Greedy function approximation: A gradient boosting machine.}''}, \href{https://doi.org/10.1214/aos/1013203451}{\emph{The Annals of Statistics} {\bfseries 29} (2001) 1189 }.

\bibitem{bib:bdt_vs_nn}
D.~McElfresh et~al., {``When do neural nets outperform boosted trees on tabular data?''}, \href{https://doi.org/https://arxiv.org/abs/2305.02997}{\emph{arXiv:2305.02997} (2024) }.

\bibitem{bib:calibration_of_juno}
A.~Abusleme, T.~Adam, S.~Ahmad, R.~Ahmed, S.~Aiello, M.~Akram et~al., {``{Calibration strategy of the JUNO experiment}''}, \href{https://doi.org/10.1007/JHEP03(2021)004}{\emph{Journal of high energy physics} {\bfseries 2021} (2021) 1}.

\bibitem{bib:neuro_mct}
{Gavrikov, Arsenii and Serafini, Andrea}, \emph{{NeuroMCT: Fast Monte Carlo Tuning with Generative Machine Learning in the JUNO Experiment}}. CHEP24, Oct., 2024, \href{https://doi.org/https://indico.cern.ch/event/1338689/contributions/6015882/}{https://indico.cern.ch/event/1338689/contributions/6015882/}.

\bibitem{bib:nufit}
{NuFIT 5.2}, \emph{Three-neutrino fit based on data available in {November} 2022}. 2022, \href{https://doi.org/http://www.nu-fit.org/}{http://www.nu-fit.org/}.

\bibitem{bib:xgboost}
T.~Chen and C.~Guestrin, {``Xgboost: A scalable tree boosting system''}, \href{https://doi.org/10.1145/2939672.2939785}{\emph{Proceedings of the 22nd ACM SIGKDD International Conference on Knowledge Discovery and Data Mining} (2016) 785–794}.

\bibitem{bib:optuna}
T.~Akiba, S.~Sano, T.~Yanase, T.~Ohta and M.~Koyama, {``Optuna: A next-generation hyperparameter optimization framework''}, \href{https://doi.org/10.48550/arXiv:1907.10902}{\emph{arXiv:1907.10902} (2019) }.

\bibitem{bib:relu}
V.~Nair and G.~E. Hinton, {``Rectified linear units improve restricted boltzmann machines''},  in \emph{Proceedings of the 27th ICML conference}, ICML'10, p.~807–814, 2010.

\bibitem{bib:lrelu}
A.~L. Maas et~al., {``Rectifier nonlinearities improve neural network acoustic models''}, {\emph{Proc. ICML} {\bfseries 30} (2013) 3}.

\bibitem{bib:silu}
D.~Hendrycks and K.~Gimpel, {``Gaussian error linear units {GELUs}''}, \href{https://doi.org/10.48550/arXiv.1606.08415}{\emph{arXiv:1606.08415} (2023) }.

\bibitem{bib:prelu}
K.~He, X.~Zhang, S.~Ren and J.~Sun, {``Delving deep into rectifiers: Surpassing human-level performance on imagenet classification''},  in \emph{Proceedings of the IEEE International Conference on Computer Vision (ICCV)}, December, 2015.

\bibitem{bib:adam}
D.~P. Kingma and J.~Ba, {``Adam: A method for stochastic optimization''}, \href{https://doi.org/10.48550/arXiv.1412.6980}{\emph{arXiv:1412.6980} (2017) }.

\bibitem{bib:sgd}
H.~Robbins and S.~Monro, {``{A Stochastic Approximation Method}''}, \href{https://doi.org/10.1214/aoms/1177729586}{\emph{The Annals of Mathematical Statistics} {\bfseries 22} (1951) 400 }.

\bibitem{bib:rmsprop}
T.~T. and H.~G., {``{Lecture 6.5 - RMSprop: Divide the gradient by a running average of its recent magnitude}''},  2012.

\bibitem{bib:ExpScheduler}
Z.~Li and S.~Arora, {``An exponential learning rate schedule for deep learning''}, \href{https://doi.org/10.48550/arXiv.1910.07454}{\emph{arXiv:1910.07454} (2019) }.

\bibitem{bib:CosAnnScheduler}
I.~Loshchilov and F.~Hutter, {``{SGDR: Stochastic Gradient Descent with Warm Restarts}''}, \href{https://doi.org/10.48550/arXiv.1608.03983}{\emph{arXiv:1608.03983} (2017) }.

\bibitem{bib:xavier}
X.~Glorot and Y.~Bengio, {``Understanding the difficulty of training deep feedforward neural networks''}, {\emph{Proceedings of the 13th AISTATS} (2010) 249}.

\bibitem{bib:orthogonal}
A.~M. Saxe, J.~L. McClelland and S.~Ganguli, {``Exact solutions to the nonlinear dynamics of learning in deep linear neural networks''}, \href{https://doi.org/10.48550/arXiv.1312.6120}{\emph{arXiv:1312.6120} (2014) }.

\bibitem{bib:batchnorm}
S.~Ioffe and C.~Szegedy, {``Batch normalization: Accelerating deep network training by reducing internal covariate shift''}, \href{https://doi.org/10.48550/arXiv:1502.03167}{\emph{arXiv:1502.03167} (2015) }.

\bibitem{bib:activation_functions}
J.~Lederer, {``Activation functions in artificial neural networks: A systematic overview''}, \href{https://doi.org/10.48550/arXiv.2101.09957}{\emph{arXiv:2101.09957} (2021) }.

\bibitem{bib:tpe}
J.~Bergstra, R.~Bardenet, Y.~Bengio and B.~K{\'e}gl, {``Algorithms for hyper-parameter optimization''}, {\emph{Advances in neural information processing systems} {\bfseries 24} (2011) }.

\bibitem{bib:torch}
A.~Paszke et~al., {``Pytorch: An imperative style, high-performance deep learning library''}, \href{https://doi.org/10.48550/arXiv:1912.01703}{\emph{arXiv:1912.01703} (2019) }.

\bibitem{bib:logloss}
I.~J. Good, {``Rational decisions''}, \href{https://doi.org/10.1111/j.2517-6161.1952.tb00104.x}{\emph{Journal of the Royal Statistical Society: Series B (Methodological)} {\bfseries 14} (1952) 107}.

\bibitem{bib:pdp}
B.~M. Greenwell, B.~C. Boehmke and A.~J. McCarthy, {``A simple and effective model-based variable importance measure''}, \href{https://doi.org/10.48550/arXiv:1805.04755}{\emph{arXiv:1805.04755} (2018) }.

\bibitem{bib:shap}
L.~Shapley, E.~Artin and M.~Morse, {``Quota solutions op n-person games''}, .

\bibitem{bib:shap-implementation}
S.~M. Lundberg and S.-I. Lee, {``A unified approach to interpreting model predictions''}, {\emph{Advances in neural information processing systems} {\bfseries 30} (2017) }.

\bibitem{bib:shap-tree-nature}
S.~M. Lundberg et~al., {``From local explanations to global understanding with explainable ai for trees''}, \href{https://doi.org/10.1038/s42256-019-0138-9}{\emph{Nature machine intelligence} {\bfseries 2} (2020) 56}.

\bibitem{bib:shap-tree}
S.~M. Lundberg, G.~G. Erion and S.-I. Lee, {``Consistent individualized feature attribution for tree ensembles''}, \href{https://doi.org/10.48550/arXiv.1802.03888}{\emph{Methods} {\bfseries 5} (2018) 25}.

\bibitem{bib:perm_imp}
L.~Breiman, {``Random forests''}, \href{https://doi.org/10.1016/S0167-9473(01)00065-2}{\emph{Machine learning} {\bfseries 45} (2001) 5}.

\bibitem{bib:mc1}
A.~P. Dawid, {``The well-calibrated bayesian''}, {\emph{Journal of the American Statistical Association} {\bfseries 77} (1982) 605}.

\bibitem{bib:calibration_of_mnn}
C.~Guo, G.~Pleiss, Y.~Sun and K.~Q. Weinberger, {``On calibration of modern neural networks''}, {\emph{Proceedings of the 34th International Conference on Machine Learning} {\bfseries 70} (2017) 1321}.

\bibitem{bib:ml_unc_review}
E.~H{\"u}llermeier and W.~Waegeman, {``Aleatoric and epistemic uncertainty in machine learning: An introduction to concepts and methods''}, \href{https://doi.org/https://doi.org/10.1007/s10994-021-05946-3}{\emph{Machine learning} {\bfseries 110} (2021) 457}.

\bibitem{bib:ml_unc_methods}
M.~Abdar et~al., {``A review of uncertainty quantification in deep learning: Techniques, applications and challenges''}, \href{https://doi.org/https://doi.org/10.1016/j.inffus.2021.05.008}{\emph{Information Fusion} {\bfseries 76} (2021) 243}.

\bibitem{bib:mc_dropout}
Y.~Gal and Z.~Ghahramani, {``Dropout as a bayesian approximation: Representing model uncertainty in deep learning''}, {\emph{Proceedings of The 33rd International Conference on Machine Learning} {\bfseries 48} (2016) 1050}.

\bibitem{bib:dropout}
N.~Srivastava, G.~Hinton, A.~Krizhevsky, I.~Sutskever and R.~Salakhutdinov, {``Dropout: A simple way to prevent neural networks from overfitting''}, {\emph{Journal of Machine Learning Research} {\bfseries 15} (2014) 1929}.

\end{thebibliography}\endgroup

\end{document}